\documentclass[usenatbib]{mn2e}

\usepackage{graphicx}
\usepackage{aasabbrev}

\renewcommand{\vec}[1]{\bmath{#1}}

\newcommand{\nd}[1]{\ensuremath{\tilde{\vec #1}}}
\newcommand{\nds}[1]{\ensuremath{\tilde{#1}}}

\newcommand{\kesc}{\ensuremath{\alpha}}
\newcommand{\pesc}{\ensuremath{\beta}}
\newcommand{\ctsc}{\ensuremath{\gamma}}
\newcommand{\grav}{\ensuremath{G}}
\newcommand{\sbc}{\ensuremath{Sbc}}
\newcommand{\g}{\ensuremath{G}}
\newcommand{\ieff}{\ensuremath{I_{\rm eff}}}
\newcommand{\reff}{\ensuremath{r_{\rm eff}}}
\newcommand{\rnfw}{\ensuremath{r_{\rm  NFW}}}
\newcommand{\rhonfw}{\ensuremath{\rho_{\rm  NFW}}}
\newcommand{\req}{\ensuremath{r_{\rm eq}}}
\newcommand{\meq}{\ensuremath{M_{\rm eq}}}
\newcommand{\meff}{\ensuremath{M_{\rm eff}}}
\newcommand{\teq}{\ensuremath{t_{\rm eq}}}
\newcommand{\mb}{\ensuremath{M_{\rm bar}}}
\newcommand{\rhob}{\ensuremath{\rho_{\rm bar}}}
\newcommand{\phib}{\ensuremath{\Phi_{\rm bar}}}
\newcommand{\md}{\ensuremath{M_{\rm DM}}}
\newcommand{\rhod}{\ensuremath{\rho_{\rm DM}}}
\newcommand{\phid}{\ensuremath{\Phi_{\rm DM}}}
\newcommand{\rhot}{\ensuremath{\rho_{\rm T}}}

\title[On Galaxies and Homology]{On Galaxies and Homology}

\author[G. S. Novak et al.]
{Gregory S. Novak,$^{1,2,3}$\thanks{E-mail: greg.novak@obspm.fr (GSN);
    pjonsson@cfa.harvard.edu (PJ); joel@scipp.ucsc.edu (JRP); 
    tcox@obs.carnegiescience.edu (TJC);  dekel@phys.huji.ac.il (AD)}, 
Patrik Jonsson,$^{4,5}$ Joel R. Primack,$^{5}$ 
\newauthor Thomas J. Cox,$^{6}$ and Avishai Dekel,$^{7}$\\
$^{1}$Observatoire de Paris, LERMA, CNRS, 61 Av de l'Observatoire, 75014, Paris, France \\
$^{2}$Department of Astrophysical Sciences, Princeton University,
Princeton NJ, 08544 \\ 
$^{3}$UCO/Lick Observatories, University of California, 
1156 High Street, Santa Cruz, CA 95064\\
$^{4}$Harvard-Smithsonian Center for Astrophysics, 
60 Garden Street, Cambridge, MA 02138 \\
$^{5}$Department of Physics, University of California, 
1156 High Street, Santa Cruz, CA 95064 \\
$^{6}$Carnegie Observatories, 813 Santa Barbara Street, Pasadena, CA 91101 \\
$^{7}$Racah institute of Physics, The Hebrew University, Jerusalem 91904, Israel}

\begin{document}

\date{Accepted 2012 May 4.  Received 2012 April 4; in original form 2010 November 4}

\pagerange{\pageref{firstpage}--\pageref{lastpage}} \pubyear{2012}

\maketitle

\label{firstpage}

\begin{abstract}  
  The definition of homology for single-component galaxies is clear,
  but for multi-component (luminous and dark matter) galaxies there is
  some ambiguity.  We attempt to clarify the situation by carefully
  separating the different concepts of homology that have been used to
  date.  We argue that the most useful definition is that a set of
  galaxies is homologous if they are the same in all respects up to a
  set of three dimensional scaling constants which may differ from one
  galaxy to the next.  Noting that we are free to choose the
  dimensional constants, we find that a set of hydrodynamic simulated
  galaxy merger remnants is significantly closer to homologous when
  the dimensional length constant is taken to be the radius containing
  equal amounts of dark and baryonic matter rather than the usual
  observationally motivated choice of the baryonic half-mass radius.
  Once the correct dimensional scaling constants are used, the stellar
  velocity dispersion 
  anisotropy is essentially the sole source of the variation in the
  kinematic structure of these simulated merger remnants.  In order to
  facilitate the use of these scaling constants to analyse observed
  galaxies, we calculated the relationship between our preferred
  dimensional scaling constants and the typical observationally
  accessible quantities.  
\end{abstract}

\section[]{Introduction}

To what extent are galaxies (or some subset of galaxies) homologous in
the sense that all individual objects are scaled copies of one
another?  The discovery of the Fundamental Plane (FP)
\citep{djorgovski:87, dressler:87} of elliptical galaxies distinctly
sharpened this question because galaxies showed a regularity that
almost, but not quite, followed that expected from the virial theorem
for systems in dynamical equilibrium.  The virial theorem holds
regardless of whether galaxies are homologous, but if each galaxy were
unique, then the FP would have a large scatter.  If all galaxies were
simply scaled copies of one another, then the FP would follow the
virial expectation.  In either of these cases, the FP would likely not
have generated the interest that it has because it would either be
less useful (in the former case) or less interesting (in the latter
case).  Actual galaxies seem to be in between these two possibilities:
they are quite regular in structure, but do not quite follow the
simple virial expectation.  An understanding of the origin of the FP
requires at least one ingredient aside from the virial theorem, and
the search for that missing ingredient has been the focus of a great
number of observational and theoretical studies.

Homology means that all galaxies are somehow scaled copies of one
another.  Obviously the property of homology applies only to a set of
objects, not to individual galaxies.  If galaxies were made up of only
one type of matter, then there is no ambiguity in the definition of homology.
However, they are in fact made up of dark matter, luminous matter, and
gas, the behaviour of each component is dominated by different physics,
and the components are only weakly coupled.  For such systems, the
proper definition of homology is not clear.  Several informal
definitions seem to be in common use; for single component systems,
all of these definitions are equivalent.  Not so for multi-component
galaxies.

We formalize the definitions currently in use and discuss which
definitions are stronger or weaker in the sense that the galaxies that
fit one definition are or are not a subset of those that fit another
definition.  Then we apply these ideas to galaxy remnants produced in
numerical simulations.  Finally, we discuss how these ideas may be
applied to observed galaxies.

It is obviously the case that homology can only apply to some subset
of galaxies (ellipticals and spirals are quite distinct).  It has also
been established that even for restricted sets of galaxies, homology
can only be approximate because, for example, the shape of the surface
brightness as a function of radius (parametrized by the S\'ersic
parameter $n$) correlates with galaxy luminosity \citep{prugniel:97}.
However, elliptical galaxies are still quite {\em close} to being
homologous--close enough that the concept is still very useful.  

In Section \ref{sec:theory} we show that homology is closely related
to the choice of units for each galaxy: homologous galaxies are all
the same up to a choice of units, where the units are allowed to
change from one galaxy to the next.  For multi-component galaxies,
there is an ambiguity in the definition of homology that we may be
able to exploit: perhaps the baryonic half-light radius is not the
right choice to make all galaxies look the same.  In Section
\ref{sec:previous-work} we review recent observational and theoretical
work on the subject of galaxy structure and homology.  In Section
\ref{sec:simulations} we use numerical simulations of galaxy mergers
to show that there is indeed a better choice of dimensional scaling
constants, at least for the set of simulations under consideration.
Unfortunately our preferred choice of scaling constants are not
directly observationally accessible, so in Section
\ref{sec:observations} we work out the relationship between our
preferred definition of the scaling constants and the conventional
choices for plausible baryonic and dark matter distributions.
Finally, in Section \ref{sec:conclusions} we summarize our
conclusions.

\section[]{Theoretical Preliminaries}
\label{sec:theory}
When discussing single-component models, it is fairly clear how to
make the definition of homology precise.  Such a galaxy is fully described by its
distribution function $f(\vec x, \vec v)$ giving the density of
stars with the given position $\vec x$ and velocity $\vec v$.  A set
of galaxies is homologous if there exists a `master' distribution
function $F$ where the distribution function of a specific galaxy $f$
is obtainable from $F$ via:
\begin{equation}
  f(\vec x, \vec v) = \frac{M}{R^3V^3} F(\vec x/R, \vec v/V)
\label{eq:nondim-df}
\end{equation}
where $R$, $V$, and $M$ are arbitrary scale constants, different for
each galaxy.  If one knows the universal distribution function, three
numbers suffice to completely describe a galaxy.  This will be called
\textit{scaling homology}.  

Scaling homology is closely related to systems of units.  It means
that all galaxies are the same if one works with the correct system of
units, where the units are allowed to change from galaxy to galaxy.

It is very important to recognize at this point that these scaling
constants are {\em not} necessarily equal to the half-mass radius, the
velocity dispersion, and the total galaxy mass.  For single component
galaxies, these choices for the scaling constants are essentially the
only reasonable ones: if galaxies are homologous, then other choices
will end up being simple functions of the half-mass radius, etc.
However, for multi-component galaxies, our job is to take a set of
galaxies and {\em infer} the definitions for $R$, $V$, and $M$ that
make all of their distribution functions derivable from a single
`master' distribution function.  If no such definition exists, then
the galaxies are not scaling-homologous.  Section
\ref{sec:simulations} is devoted to carrying out this analysis for a
set of simulated galaxy merger remnants.

Perhaps the most common method of defining homology is to define
structure constants relating the true values of a galaxy's mass and
luminosity to `virial estimates' obtained using scaling constants.
From \citet{bender:92}:
\begin{equation}
  L = c_1 \ieff \reff^2
\end{equation}
\begin{equation}
  M = c_2 \sigma_0^2 \reff
  \label{eq-structure-constant}
\end{equation}
where $L$ is the galaxy's luminosity, $M$ is the mass, \reff\ is the
projected half-light radius, \ieff\ is the mean surface brightness
inside \reff, $\sigma_0$ is the projected aperture velocity
dispersion, and $c_1$ and $c_2$ are structure constants defined by
these two equations.  The important point is that $M$ is the true mass
of the galaxy while $\sigma_0$ and \reff\ are scale factors combined
in such a way as to dimensionally produce a mass (up to physical
constants).  Therefore variation of $c_2$
from galaxy to galaxy indicates that mass profiles of the two galaxies
differ to produce a different relationship between the virial
estimator and the actual mass.  A set of galaxies is said to be
homologous if these structure constants are the same for all galaxies.
We denote this definition \textit{structure-constant homology}.

It is easy to show that for single-component systems,
structure-constant homology and scaling homology are equivalent.
Putting the two definitions into the same language, we have:
\begin{equation}
  \rho(\vec x) = \int f(\vec x,\vec v) \, d^3\vec v
\end{equation}
the total kinetic energy is 
\begin{equation}
  \mbox{KE} = \frac{1}{2} \int |\vec v|^2 f(\vec x,\vec v) \, d^3\vec
  x \, d^3\vec v
\end{equation}
and the total potential energy is: 
\begin{eqnarray}
  \mbox{PE} & = & - \frac{\grav}{2} 
  \int \frac{ \rho(\vec x) \rho(\vec y) 
\, d^3\vec x \, d^3\vec y} 
{|\vec x - \vec y|}  \nonumber \\
& = & - \frac{\grav}{2} 
  \int \frac{f(\vec x,\vec v) f(\vec y,\vec w)  
\, d^3\vec x \, d^3\vec y \, d^3\vec v \, d^3\vec w} {|\vec x - \vec y|}  
\end{eqnarray}
where $\grav$ is Newton's gravitational constant.
The scalar virial theorem is then:
\begin{eqnarray}
& & \int |\vec v|^2 f(\vec x, \vec v) \, d^3\vec x \, d^3\vec v \nonumber \\
& & = \frac{\grav }{2} \int \frac{f(\vec x,\vec v) f(\vec y,\vec w)  
\, d^3\vec x \, d^3\vec y \, d^3\vec v \, d^3\vec w} {|\vec x - \vec y|} \, .
\end{eqnarray}
Let us define three arbitrary scaling constants $R$, $V$, and $M$, and
dimensionless coordinates $\nd{x}=\vec x/R$ and $\nd{v}=\vec
v/V$.  
\begin{eqnarray}
& & R^3 V^5 \int |\nd{v}|^2 f(\vec x, \vec v) \, d^3\nd{x} \, d^3\nd{v}
\nonumber \\
& & = \frac{\grav R^5 V^6}{2} \int \frac{f(\vec x,\vec v) f(\vec y,\vec w)  
\, d^3\nd{x} \, d^3\nd{y} \, d^3\nd{v} \, d^3\nd{w}}
{|\nd{x} - \nd{y}|}
\end{eqnarray}
Using the definition of the dimensionless distribution function in
Equation \ref{eq:nondim-df}, the virial theorem becomes:
\begin{eqnarray}
& & M V^2  \int |\nd{v}|^2 F(\nd{x}, \nd{v}) \, d^3\nd{x} \, d^3\nd{v}
\nonumber \\ 
& & =  \frac{\grav M^2}{2R} \int \frac{F(\nd{x},\nd{v}) F(\nd{y},\nd{w})
\, d^3\nd{x} \, d^3\nd{y} \, d^3\nd{v} \, d^3\nd{w}}
{|\nd{x} - \nd{y}|}
\end{eqnarray}
Therefore we define the structure constants
\begin{eqnarray}
\kesc &=&  \int |\nd{v}|^2 F(\nd{x}, \nd{v}) \, d^3\nd{x} \, d^3\nd{v}
\nonumber \\ 
&=& \frac{1}{M} \int |\vec v/V|^2 f(\vec x, \vec v) \, 
d^3\vec x \, d^3\vec v 
\end{eqnarray}
and
\begin{eqnarray}
\pesc &=&
\int \frac{F(\nd x,\nd v) F(\nd y,\nd w) \, d^3\nd{x} \, d^3\nd{y} \, d^3\nd{v} \, d^3\nd{w}}
{|\nd{x} - \nd{y}|} \nonumber \\ 
&=& \frac{R}{M} \int \frac{f(\vec x,\vec v) f(\vec y,\vec w) \,
  d^3\vec x \, d^3\vec y \, d^3\vec v \, d^3\vec w}
{|\vec x - \vec y|}
\end{eqnarray}
With these definitions, the virial theorem is simply:
\begin{equation}
\kesc V^2 = \pesc \frac{\grav M}{2R}
\end{equation}
and comparing to equation \ref{eq-structure-constant} reveals that:
\begin{equation}
  c_2 = 
\left(\frac{2 \alpha}{ \grav \beta} \right)
\left(\frac{\mb}{M} \right)
\left(\frac{R}{\reff} \right)
\left(\frac{V}{\sigma_0} \right)^2
\end{equation}
where \mb\ refers to the total baryonic mass of the galaxy.  

\subsection{Two Component Galaxies}
Describing a galaxy that contains both stellar and dark matter
requires specifying two distribution functions, f and g, with
corresponding dimensionless functions F and G.

Why do we use separate distribution functions for baryonic matter and
dark matter, but we do not separate the different components of
baryonic matter into young stars, old stars, disc stars, etc?  The
different stellar components have different formation histories and
the moments of the distribution function that appear in the Jeans
equations will be different for the different stellar components.
Here we attempt to strike balance between accuracy and simplicity: our
description of the system should be complex enough to be accurate, but
simple enough to be informative.  Hence we separate baryonic matter
from dark matter because the two types of matter obey different
physics and have {\em very} different histories.  By including all
baryonic matter in a single distribution function, we sacrifice some
accuracy in order to make the description simple enough to be easily
understood and manipulated.  

Allowing for a second component, the virial theorem takes the form:
\begin{eqnarray}
& & \frac{2 R V^2}{\grav M}  \left( \int |\nd{v}|^2 F(\nd{x}, \nd{v}) \, d^3\nd{x} \, d^3\nd{v}
+ \int |\nd{v}|^2 G(\nd{x}, \nd{v}) \, d^3\nd{x} \, d^3\nd{v} \right)
\nonumber \\
& & = \int \frac{F(\nd{x},\nd{v}) F(\nd{y},\nd{w})}{|\nd{x} - \nd{y}|} 
\, d^3\nd x \, d^3\nd y \, d^3\nd v \, d^3\nd w \nonumber \\
& & + 2 \int \frac{F(\nd{x},\nd{v}) G(\nd{y},\nd{w})}{|\nd{x} - \nd{y}|} 
\, d^3\nd x \, d^3\nd y \, d^3\nd v \, d^3\nd w \nonumber \\
& &  + \int \frac{G(\nd{x},\nd{v}) G(\nd{y},\nd{w})}{|\nd{x} - \nd{y}|}  
\, d^3\nd x \, d^3\nd y \, d^3\nd v \, d^3\nd w \, .
\end{eqnarray}
We define 
\begin{eqnarray}
  \ctsc & = &
2 \int \frac{F(\nd{x},\nd{v}) G(\nd{y},\nd{w}) 
\, d^3\nd{x} \, d^3\nd{y} \, d^3\nd{v} \, d^3\nd{w}}
{|\nd{x} - \nd{y}|} \nonumber \\
& = & \frac{2 R}{M} \int \frac{f(\vec x,\vec v) g(\vec y,\vec w) \, 
d^3\vec x \, d^3\vec y \, d^3\vec v \, d^3\vec w}
{|\vec x - \vec y|}
\end{eqnarray}
so that at last we have the virial theorem in the form:
\begin{equation}
(\kesc_f + \kesc_g)V^2 = (\pesc_f + \pesc_g + \ctsc)\frac{\grav M}{2R} \, .
\end{equation}

From these expressions it is clear that for multi-component galaxies:
\begin{itemize}
\item Scaling homology implies structure-constant homology.
\item Structure-constant homology does not imply scaling-homology
  because structure-constant homology only requires a small number of
  constraints on the definite integrals defining \kesc, \pesc, and
  \ctsc.
\item A set of galaxies being structure-constant homologous does not
  imply that each component is separately structure-constant
  homologous.  Structure-constant-homologous galaxies can be built
  from non-homologous components if the change in $\kesc_f$
  compensates for the change in $\kesc_g$ and similarly for \pesc.
\item The components of galaxies being structure-constant-homologous
  does not imply that the galaxies are structure-constant-homologous.
\item The structure constants for multi-component galaxies are not
  simply the sum of the structure constants of the individual
  components owing to the cross term \ctsc.
\end{itemize}

Finally, similar considerations for scaling-homology yield:
\begin{itemize}
\item Scaling-homology of galaxies implies scaling-homology of the
  components.
\item Scaling-homology of the components does not imply
  scaling-homology of galaxies.  If a set of galaxies is composed of
  scaling-homologous components, then the set of multi-component
  galaxies will be scaling-homologous only if the ratios of the
  scaling constants for each component are the same for all of the
  galaxies in the set.  That is, $R_f/R_g$ must be the same for all of
  the galaxies.
\end{itemize}

One may wonder whether the dark and luminous components should be
allowed their own scaling constants.  This idea is problematic because
scaling the components separately disturbs the dynamical equilibrium
of the galaxy.  Consider a galaxy in dynamical equilibrium where the
baryonic scale radius is allowed to shrink to zero, making the baryons
into a point mass.  The dark matter far outside the original baryonic
scale radius will not be significantly affected by this operation.
However, the dark matter near the centre (inside the original baryonic
scale radius) will have more mass enclosed after the change.  The new
galaxy will obviously be out of dynamical equilibrium.

We have seen from the above considerations that scaling homology is a
stronger concept that structure-constant homology in the sense that
scaling homology implies structure-constant homology, but not vice
versa.  For two-component galaxies, the two components can undergo
compensating changes to leave the structure constants unchanged.  This
does not seem to be a desirable property for a concept like homology.
Finally, there is a certain theoretical elegance to the viewing
homology as closely connected to the choice of the system of units---a
set of galaxies is scaling homologous if all of the galaxies are the
same when using the correct system of units.  For all of these reasons
we advocate that homology be taken to mean scaling homology.
Structure-constant homology should be clearly denoted as a different
concept.

\subsection{Tilt of the Fundamental Plane}
There are three independent ways to generate the tilt in the
fundamental plane.  Any or all of them may be at work.

The first is a systematic change in the stellar mass-to-light ratio
with the mass of the host galaxy via trends in the age, metallicity,
or initial mass function of the stellar population.  In this case the stellar and
dark matter distribution functions are scaling (and therefore
structure-constant) homologous because the masses of stars and dark
matter components are unaffected.  

The second is a systematic change in the dark matter fraction near the
centre of the galaxy.  This is a mild form of non-homology which is
achieved by changing the ratio of the mass scaling constants $\mb/\md$
with galaxy mass, where \md\ is the total dark matter mass of the
galaxy.  Overall scaling- and structure-constant homology are
violated, but the baryonic components by themselves may be nearly
scaling- and structure-constant-homologous.  Furthermore, as long as
the stellar and dark matter remain nearly decoupled due to their very
different half mass radii, then changing the dark matter fraction
should not greatly disturb the dynamical equilibrium of the stellar
material.  In this case the stellar population mass-to-light ratio is
identical for all galaxies, but the total mass-to-light (both for the
entire galaxy and within any radius) ratio changes systematically with
mass.

The third possibility is that homology is flagrantly violated and the
baryonic components of galaxies are neither scaling nor
structure-constant homologous.  Even if the stellar population
mass-to-light ratio is always the same, there is still total freedom
in the total mass-to-light ratio owing to the total freedom in the
choice of dark matter profile for each galaxy.  It may change or be
constant as a function of radius, and the total mass-to-light ratio
may change or be constant for the entire set of galaxies.

A great source of confusion is that the first and second possibilities
are very different theoretically but rather similar 
observationally.  Everyone agrees that the first possibility involves
homology and the third possibility involves non-homology.  However,
the second possibility is {\em observationally} very similar to the
first possibility but {\em theoretically} very similar to the third
possibility.  This unfortunate circumstance makes it difficult to come
to firm conclusions.  

\subsection{Variable dark matter fraction and equilibrium}
There are two contributions to a total mass-to-light ratio: one from
the characteristics of the stellar population and one due to the
accompanying dark matter.  If the mass-to-light ratio changes with
galaxy mass because of a changing stellar population, then the FP will
be tilted but the mass distribution may still be homologous.  However,
if the mass-to-light ratio changes due to changing dark matter
fraction, then the galaxies {\em must} be non-homologous.

Consider the spherically symmetric Jeans equation \citep[e.g.][]{BT}:
\begin{equation}
  \frac{d \ln\nu}{d \ln r} + \frac{d \ln\sigma_r^2}{d \ln r} 
+ 2\beta_{\rm a} + \frac{1}{\sigma_r^2}\frac{d\Phi}{d \ln r} = 0
\end{equation}
where $\nu$ is the number density of tracer particles, $\sigma_r$ is
the radial component of the velocity dispersion, $\Phi$ is the total
gravitational potential, $\beta_{\rm a} = 1-\sigma_\theta^2/\sigma_r^2$ is the
anisotropy parameter, and $\sigma_\theta$ is the tangential component
of the velocity dispersion.

The matter density is provided by two components, \rhob\ and
\rhod\ for baryons and dark matter.  The Poisson equation is linear
so the solutions obey the superposition principle.  
Putting this into the Jeans equation results in:
\begin{equation}
  \frac{d \ln\nu}{d \ln r} + \frac{d \ln\sigma_r^2}{d \ln r}  +
  2\beta_{\rm a} + 
\frac{1}{\sigma_r^2}
\left(\frac{d\phib}{d \ln r} + b\frac{d\phid}{d \ln r}\right)  = 0
\end{equation}
where \phib\ and \phid\ refer to the potential due to the baryonic and
dark matter respectively and the constant $b$ parametrizes the dark
matter fraction.  If a galaxy is in equilibrium for any value of $b$,
then {\em any} change to $b$ {\em must} be compensated by a change in
the anisotropy $\beta_{\rm a}$ or a change in the radial fall-off of density
or velocity dispersion.

Variations in the dark matter fraction and non-homology are the same
concept.  If the tilt in the FP is {\em not} due to systematic changes
in the stellar population mass-to-light ratio, then it {\em must} be
due to non-homology.  There is no separate case where `only' the
dark matter fraction and therefore the observed mass-to-light ratios
are changing.

\subsection{What constitutes non-homology?}

Given the somewhat confused use of the word homology in the present
literature, it is perhaps easier to focus on what {\em non-homology}
means.  As discussed above, one way galaxies could be non-homologous
is by having baryonic profiles that change shape from galaxy to
galaxy.  The fact that not all galaxies have the same S\'ersic index
already tells us that the population of galaxies cannot be perfectly
homologous.  If the dark and luminous components scale separately from
each other so that different galaxies have different amounts of dark
matter within some radius, then the galaxies must also be considered
non-homologous.  

Finally, recall that since structure-constant homology only places a
few constraints on the definite integrals defining $\alpha$, $\beta$,
and $\gamma$, the luminous and dark components of a set of galaxies
can undergo compensating changes so that the galaxies {\em would} be
deemed structure-constant homologous, but would {\em not} be scaling
homologous.  We use this observation to argue that scaling homology is
the more useful concept and the concept of structure-constant homology
should be avoided.  

\section[]{Recent Work}
\label{sec:previous-work}

Armed with a precise definition of the different flavors of homology
we highlight a few recent observational and theoretical results.  We
wish to give a brief overview of the state of the field, but
unfortunately none of these studies will be directly relevant to the
present one because they all use the half-light radius as the relevant
length scale.  In section \ref{sec:simulations} we will show that for
a large set of hydrodynamic simulations, we obtain more instructive
results by using $\req$, the radius where the enclosed dark matter and
luminous matter are equal.  Simulated merger remnants look
significantly more homologous when using $\req$ than when using
$\reff$.  

Thus far we have been concerned only with theoretical galaxies: all
quantities of interest can be measured to arbitrary precision.
However, the only observationally accessible quantities are projected
ones.  The measured mean surface brightness within the effective
radius involves an integral along the light of sight and therefore
implicitly mixes information about the entire mass/light profile at
all radii.  Furthermore, both dark matter and stellar populations
contribute to the measured mass-to-light ratios of galaxies.
Variations of mass-to-light ratios due to one or the other of these
possibilities are very different theoretically, but observationally
similar.  

One may ask if we could take the results of an observational or
theoretical study discussed below, take a guess at the scaling of the
density profile between $\req$ and $\reff$, and then directly compare
the results of previous studies to our own.  Unfortunately, carrying
out such a project would tell us something about our assumed scaling
between $\reff$ and $\req$, but it would not tell whether other
authors would reach different conclusions if they use $\req$ instead
of $\reff$.  In order to answer the latter question, we need access to
the full density profiles as a function of radius in the case of
numerical studies.  We unfortunately must wait for other groups to
analyse their simulations in terms of $\req$ in order to see if our
findings hold true beyond our present set of simulations.

\subsection{Observational Studies}
There has been some success in recent years indicating that stellar
population models have progressed to the point where it is possible to
measure stellar mass-to-light ratios from spectra
\citep[e.g.][]{graves:10-mass-to-light}.  The usual assumption is that
the initial mass function is universal, but recent observations have
cast doubt on that idea \citep{vdokkum:11, cappellari:12-abridged}.  Furthermore, in spite
of the many codes available to estimate stellar masses
\citep[e.g.][]{bell:03, kauffmann:03, gallazzi:05, blanton:07}, 
many are based on a rather small number of libraries of stellar spectra 
\citep[e.g.][]{bruzual:03}.  Thus one may be concerned about
covariance between the different estimates of stellar mass because
their fundamental assumptions are similar.  If this is the case, then
agreement between the different estimates indicates that their
assumptions are similar, not that the stellar masses have actually
been accurately measured.

\citet{cappellari:06} used \citet{schwarzschild:79} modeling of SAURON
\citep{bacon:01} early-type
galaxies to show that the so-called virial mass estimator ($5 \reff
\sigma^2 / G$) is an excellent estimator of the dynamical mass within
one effective radius.  The SAURON survey also has stellar spectra over
the whole galaxy image out to approximately one effective radius, and is thus excellently
placed to use stellar population synthesis to finally break the
degeneracy between dark matter fraction and stellar population
mass-to-light ratio.  Their Figure 17 unfortunately does not permit a
single firm conclusion.  For dynamical mass-to-light ratios below 2.8,
there is a linear relationship between the stellar mass-to-light ratio
and the dynamical mass-to-light ratio, indicating that the dark matter
fraction does not change.  For larger mass-to-light ratios, there is
essentially a single stellar mass-to-light ratio, independent of the
dynamical mass-to-light ratio, indicating that either the initial mass
function or the the dark matter fraction {\em does} change.  

\citet{bolton:08-fp} and \citet{koopmans:09} used strong gravitational
lenses to argue that if one replaces the luminosity surface density in
the FP with the mass surface density, one gets an untilted fundamental
mass plane that is in full agreement with the virial expectation.
Therefore the tilt in the FP must be due to variations in
mass-to-light ratio.  However, there is little guidance from strong
gravitational lensing about whether the total mass-to-light variation
is due to changing dark matter fractions, changing stellar
populations, or a changing initial mass function.  

\citet{taylor:10} used SDSS data to compare stellar mass and dynamical
mass estimates for a set of galaxies.  They found that the two were
related, but non-linearly, and that the non-linearity could be removed
by defining a `structure-corrected' dynamical mass that takes the
S\'ersic parameter $n$ of the galactic light into account.  Thus they
conclude that elliptical galaxies are not homologous.
\citet{taylor:10} made the conventional choice of scaling different
galaxies by their half-light radii and central velocity dispersion in
order to evaluate the question of whether or not they are homologous.

\citet{graves:10-mass-to-light} study stacked SDSS spectra
of early-type galaxies as a function of their position in the
Fundamental Plane.  They found that the variation in the dynamical
mass-to-light ratios was too large to be explained by variations in
the stellar mass-to-light ratios alone.  This indicates varying dark
matter fractions and is interpreted as non-homology in the galaxy
population.

\citet{auger:10} used multi-band Hubble Space Telescope images to
construct stellar masses from stellar population models for 73 massive
early-type galaxies identified as strong gravitational lenses as part
of the SLACS survey \citep{czoske:08}.  Combining this information
with velocity dispersions from the Sloan Digital Sky Survey (SDSS)
\citep{york:00-abridged}, they concluded that the central dark matter
fractions of massive early-type galaxies increase systematically with
galaxy mass if one assumes a universal initial mass function.  They
found that the stellar mass-to-light ratio was essentially constant
with galaxy mass.  \citet{barnabe:11} carried out dynamical modeling
of a subset of these galaxies using ground-based integral-field-unit
data and reached the same conclusion.

\citet{grillo:10-sdss} analyzed SDSS velocity dispersions along with
stellar masses obtained from the JHU/MPA value-added galaxy 
catalog.\footnote{http://www.mpa-garching.mpg.de/SDSS/}  Dynamical
considerations led them to conclude that the ratio of dynamical mass
to dark mass was constant and they argued that the tilt of the FP
is entirely due to stellar population effects.  
\citet{grillo:10-coma} used \citet{schwarzschild:79} models of
13 bright galaxies in the Coma cluster constructed by
\citet{thomas:07,thomas:09} along with stellar masses from SDSS to
reach the same conclusion.  

After many years of effort, it seems that stellar population modeling
and dynamical modeling have progressed to the point where it is
possible to separate the effects of changing dark matter fractions
from that of changing stellar populations.  However, recent studies
have concluded both that galaxies are homologous and that they are not
homologous.  In addition, most applications of stellar population
modeling require the assumption of a universal initial mass function.
However, work by \citet{vdokkum:11} and \citet{cappellari:12-abridged} have
cast doubt on this long-held assumption.  If the initial mass function
is truly not universal, both dynamical modeling and stellar population
modeling will be more difficult.

\subsection{Theoretical Studies}

\citet{gonzalez-garcia:03} and \citet{boylan-kolchin:05} studied
dissipationless merger simulations to determine whether mergers leave
the FP intact.  \citet{gonzalez-garcia:03} started with
structure-constant homologous systems and found that their remnants
did not share the property.  The merger remnants {\em do} lie on the
FP, evidently because the two structure constants are able to undergo
compensating changes since they only appear in a multiplicative
combination in the FP relation (their Equation 6).
\citet{boylan-kolchin:05} studied a smaller number of much higher
resolution simulations and found that both structure-constant homology
and the FP were preserved by the merger process at the 10\% level.
They did, however, find a significant violation of homology for
radial-orbit mergers.

\citet{dekel:06-fp} used numerical simulations to study dissipation as
the potential origin of the FP tilt.  They assumed that all observable
galaxy properties are power-law functions of the mass of the galaxy
and the gas fraction, and then used several merger simulations with
different gas fractions and masses in order to fix the power-law
exponents and constants in their model.  They were left with a set of
constraints on the power-law exponents such that for any observed FP
tilt, one could solve for the properties of the progenitor galaxies
(e.g., gas fraction) as a function of mass.  This of course depends on
their merger simulations being good representatives of all mergers of
that mass and gas fraction.

\citet{robertson:06-fp} used a large number of dissipational
simulations of galaxy mergers, assumed a constant stellar
mass-to-light ratio and concluded that with a sufficient amount of
gaseous dissipation, merger remnants naturally produce a tilted FP.
Since they assumed a constant stellar mass-to-light ratio, this tilt
must have been caused by structural non-homology/varying dark matter
fraction.

\citet{hopkins:08-fp} performed an extensive study of dissipation and
the origin of the fundamental plane.  They seem to be concerned with
structure-constant homology, but dismiss non-homology as the source of
the FP tilt.  They assert that gas dissipation causes a varying
dark-matter fraction inside the effective radius, which in turn tilts
the FP.  As stated above, we note that a varying dark-matter fraction
is not distinct from non-homology.  However, it can perhaps be
considered to be a weak form of non-homology where only one of many
possible degrees of freedom is exercised.

\subsection{Relation to the present study}

It is conceivable that any or all of these studies would reach a
different conclusion if they were to use our definition of length,
velocity, and mass scales rather than the usual observationally
motivated definitions.  In order to make this determination, we need
more information than is available in the published papers, so we must
wait for other groups to try analyzing their simulations and
observations in terms of $\req$ rather than $\reff$. 

We hasten to add that we are {\em not} claiming that any previous
authors have made any mistake.  Other researchers have correctly
formulated and answered questions about the scaling relations of
galaxies.  Given the assumption that the relevant length scale was the
baryonic half-mass radius, these studies generally found measurable
non-homology in the galaxy population that manifested itself by a
changing central dark matter fraction.  

We are pointing out that the choice of $\reff$ as the relevant length
scale is arbitrary, and other choices may be more instructive in
understanding galaxies and galaxy merger remnants.  On the bases of
our simulations, we have found that scaling the simulations by $\req$
gives very instructive results.  We hope that other researchers will
find this terminology similarly useful.

\section[]{Application to Simulations}
\label{sec:simulations}
As already noted, the dimensional scaling constants used in the
definition of scaling-homology are {\em not} required to be the
conventional choices of half-mass radius, central velocity dispersion,
etc.  In this section we consider a set of simulated galaxy merger
remnants and search for a definition of the dimensional scaling
constants so that the dimensionless properties of the set of
remnants are as similar as possible.  

Once one has arrived at a definition for the scaling constants that
seems to work well, one must compute the value for a given galaxy.
Therefore we are restricting ourselves to simple, physically-motivated
definitions of the dimensional scaling constants, such as the baryonic
half-mass radius, the total half-mass radius, the radius where the
enclosed baryonic and dark masses are equal, etc.  

One could imagine carrying out a high-dimensional minimization where
the function to be minimized is the sum of the total deviations of the
scaled density and velocity profiles from the mean scaled profiles for
all galaxies.  The free parameters in the minimization are the three
scaling constants for each galaxy.  This would effectively tell you
whether or not there {\em exists} a definition of the scaling
constants that makes the galaxies scaling-homologous.  However, it
would {\em not} give a simple way to compute the dimensional scaling
constants for a {\em given} galaxy.  One would {\em know} the value of
the scaling constants for a galaxy in the `training' set: the
fitting procedure would provide it.  But one would not be able to
compute the dimensional constant for a {\em new} galaxy, not in the
training set.

Using numerical simulations allows us to quickly explore many possible
definitions of the scaling constants with little uncertainty in, for
example, the total galaxy mass or velocity dispersion.  The simulated
galaxies are of course imperfect stand-ins for actual galaxies, but
our hope is that working with simulated galaxies will allow us to
eliminate many of the definitions that are possible and focus on one
(or a few) that seem to work well.  Only the most successful
choices when compared to simulations will be considered including the
additional complications inherent to observational data.

There are two sets of simulations of binary mergers considered here.
The \sbc\ series consists of equal mass mergers between two identical
gas-rich progenitors on a variety of orbits.  The \g\ series consists
of mergers of galaxies with lower gas fractions and a wide range 
mass ratios using only a few different orbits.
\citet{cox:06-feedback} contains a full description of the
simulations.

We considered eight possible radius scales including stellar, dark,
and total 3D effective radii and projected effective radii, the radius
where stellar mass density and dark mass density are equal, and
finally the radius where the enclosed stellar mass and enclosed dark
mass are equal.  We evaluated thirty mass scales including the
stellar, dark, and total mass, the total of each type of mass within
each of four radii (stellar effective radius, dark effective radius,
radius where stellar and dark densities are equal, and the radius
where the enclosed stellar and dark masses are equal), and the mass
scale that sets each of the three types of mass density to one at each
of the four radii.  The last dimensional choice is either a velocity
or a time, and we tried the central 3D velocity dispersion, the
projected velocity dispersion, and dynamical times at each of the
above listed radii.  

Evaluating one `model' requires a choice of one of each of the three
dimensional constants: length, mass, and velocity/time.  This leads to
a very large number of combinations.  Fortunately, only a few choices
work better than the conventional observationally motivated choices of
baryonic half-mass radius, central velocity dispersion, and total
baryonic mass.  In this paper we plot only the conventional choices
and the best alternate definition we found: the length scale is \req,
the radius such that the enclosed dark and baryonic matter are equal,
\meq, the mass enclosed within \req, and \teq, the
dynamical time at \req.  For the dynamical time, we use 
\begin{equation}
  t_{\rm dyn} = \sqrt{\frac{3\pi}{32 G \rho}} 
\end{equation}
so that
\begin{equation}
  \teq = \sqrt{\frac{\pi^2\req^3}{8 G \meq}} \, . 
\end{equation}

To evaluate each choice of the scale constants, we plot
dimensionless density and velocity dispersion profiles for all
galaxies:
\begin{equation}
  \nds{\rho}(r) = \frac{R^3 \rho (r/R)}{M} 
\end{equation}
\begin{equation}
  \nds{\sigma}(r) = \frac{\sigma (r/R)}{V} 
\end{equation}
where $r$ is the magnitude of the radius vector.

For simplicity we use spherical apertures to convert the 3D density
and velocity profiles of the simulated galaxies into the 1D profiles
shown here.  Using ellipsoidal profiles instead does not change our
conclusions.  The velocity dispersion and rotation information is
given in spherical coordinates $r, \theta,$ and $\phi$ where $\phi$ is
the azimuthal angle and the coordinate system is defined by the shape
of the stellar remnant so that the $z$ axis corresponds to the short
axis of the remnant.  

If the simulated galaxies are perfectly scaling-homologous and we have
hit upon the correct definition of the dimensional scaling constants,
then the dimensionless density and velocity dispersion profiles will
be exactly the same for all galaxies.  Non-scaling-homologous galaxies
will lead to large variations in the profiles from galaxy to galaxy.
Therefore we plot the dimensionless profiles for {\em all} galaxies
(with some transparency in each line so that the effects of
over-plotting are minimized) and seek the definition of scaling
constants that leads to the least variation from galaxy to galaxy.  

For each simulation, we must estimate various observationally relevant
quantities from simulation snapshots.  For real galaxies, observables
such as the half-light radius, aperture velocity dispersion, and
luminosity are affected by star formation history, age gradients,
metallicity gradients, and dust.  We ignore the majority of these
complications and make simple assumptions that allow us to focus on
the essential physics that determine the structure of merger remnants.
In the future, a more detailed study using a code that allows us to
take the effects stellar populations and dust into account
\citep[e.g.][]{jonsson:06, jonsson:10} would be worthwhile.

For the moment, we assume a constant stellar mass-to-light ratio and no
dust scattering or absorption allowing for a simple relationship
between mass and light.  To estimate 3D and aperture velocity
dispersions, we perform a maximum likelihood fit of a Gaussian
distribution to the simulation particle velocities.  The stellar remnants are very close
to oblate spheroids, so to estimate the projected half-light radius we
use a ``typical'' viewing angle 45 degrees between the pole-on and
face-on views.  We fit an ellipse to the projected iso-density
contours and use the geometric mean of the semi-major and semi-minor
axes as the half-light radius.  

\begin{figure}
  \centering
  \includegraphics[width=0.9\columnwidth]{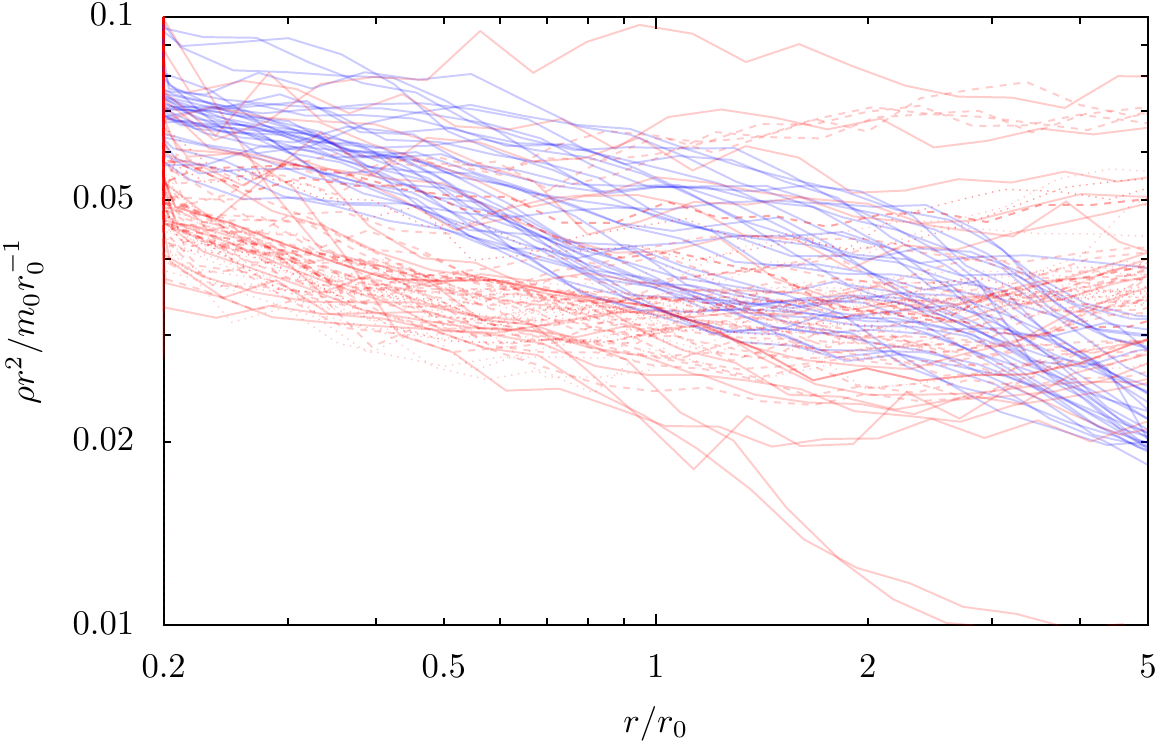}
  \includegraphics[width=0.9\columnwidth]{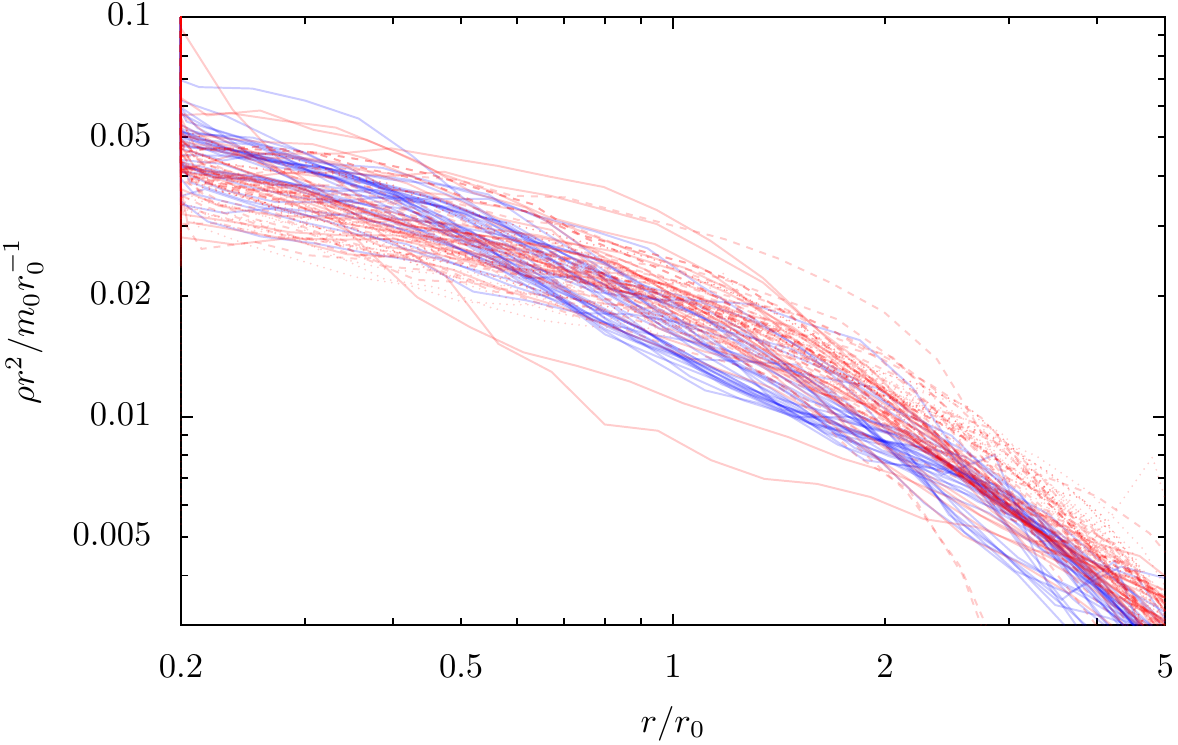}
  \includegraphics[width=0.9\columnwidth]{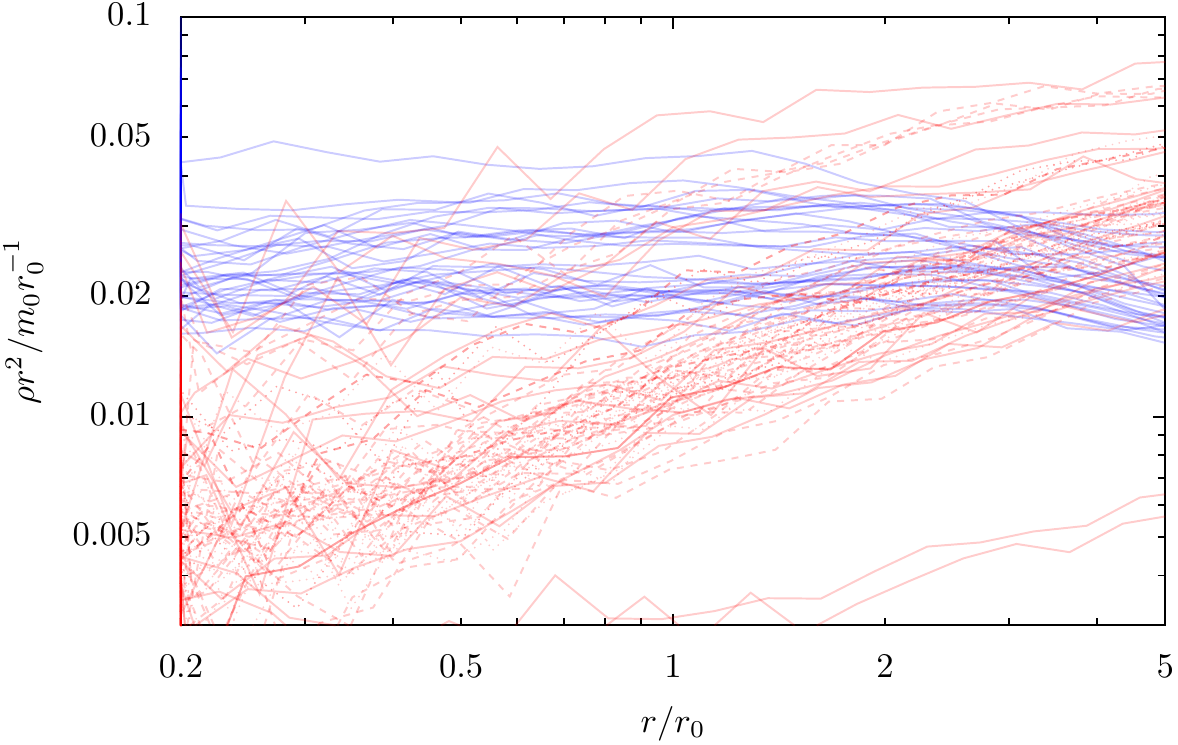}  
  \caption{The dimensionless mass density profiles for all \g\
    and \sbc\ series simulations using the typical observationally
    motivated choices of stellar half mass radius and stellar mass for
    the length and mass scales.  Blue lines are for \sbc\ simulations
    (with gas rich progenitors) while red lines are \g\ series
    simulations.  From the top, the total mass density, the stellar
    mass density, and the dark matter mass density.  All mass
    profiles have been multiplied by $r^2$ to flatten them out.  The
    stellar components of all of the simulations are strikingly
    similar.  However, the chosen mass and length scales have nothing
    to do with the dark matter, so there is significant variation
    among the dark matter profiles of the different simulations.
    There is also a systematic change in the slope of the dark matter
    profiles between \sbc\ and \g\ series simulations.  No change of
    units will remove this difference.}
  \label{fig:homology-density}
\end{figure}

Figure \ref{fig:homology-density} shows the dimensionless
density profiles for the usual choice of the stellar effective radius
for the length scale and the total stellar mass for the mass scale.

\begin{figure}
  \centering
  \includegraphics[width=0.9\columnwidth]{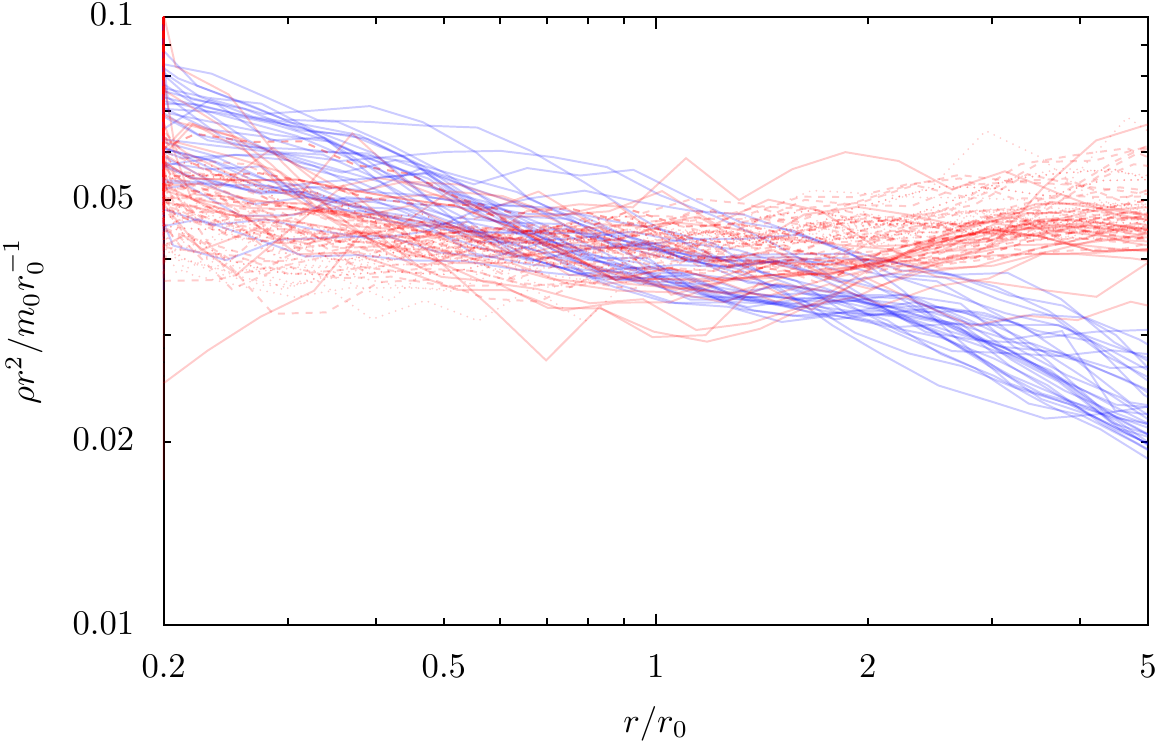}
  \includegraphics[width=0.9\columnwidth]{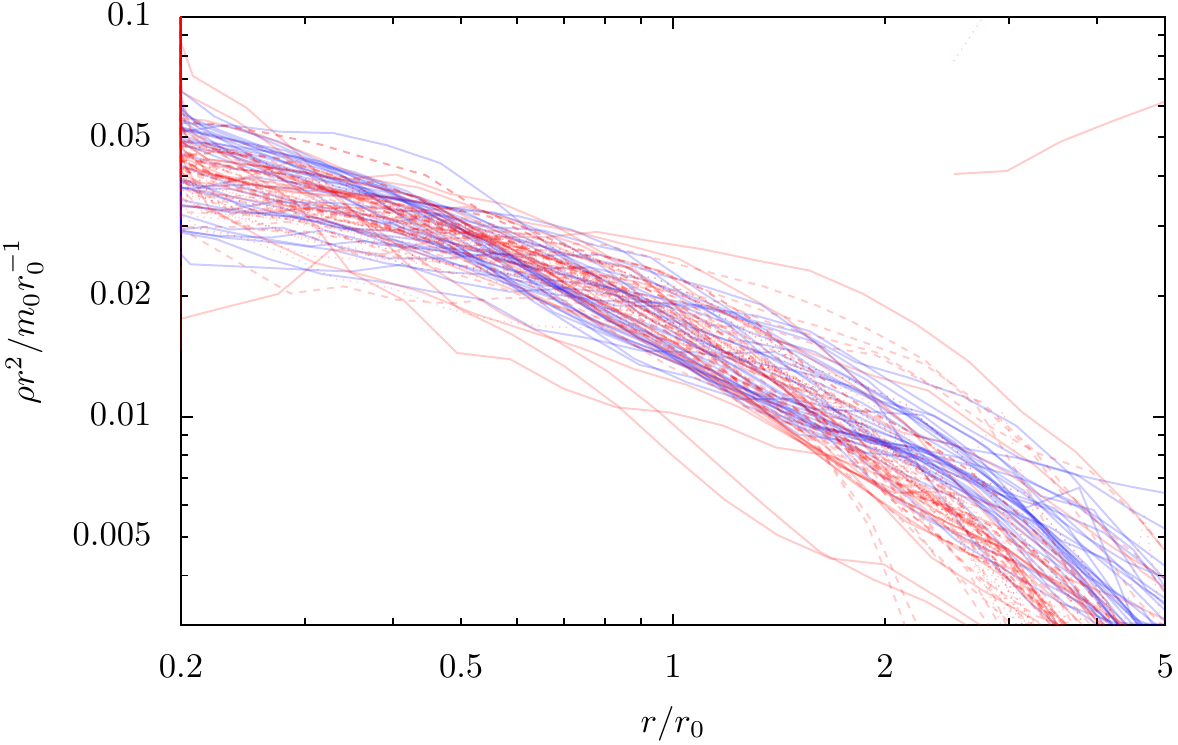}
  \includegraphics[width=0.9\columnwidth]{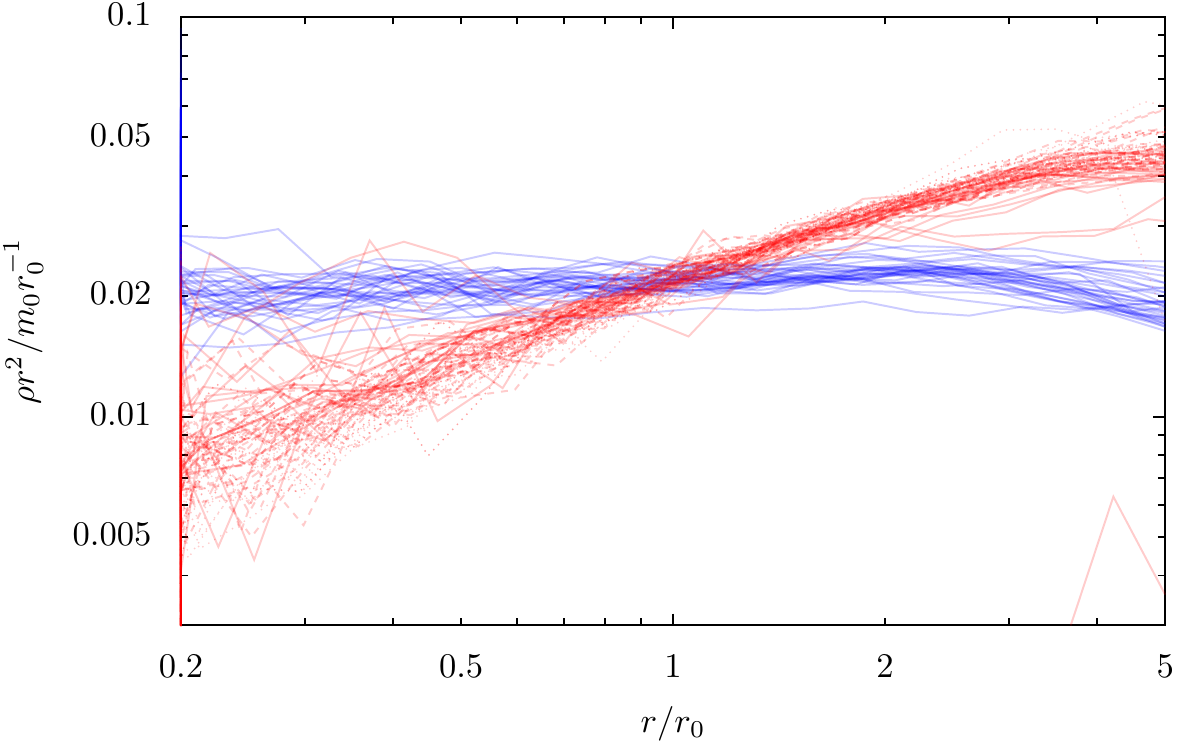}  
  \caption{The dimensionless mass density profiles for all \g\
    and \sbc\ series simulations using our preferred choice of radius
    scale of $0.437\req$, where \req\ is the 3D radius where the
    enclosed stellar mass and dark mass are equal, and the mass scale
    is the 1.54 times the stellar mass enclosed within \req.  The
    factors of 0.437 and 1.54 are chosen to ensure that these plots
    cover nearly the same physical region of the galaxy as Figure
    \ref{fig:homology-density}.  With this choice of length and mass
    scale, the merger remnants show remarkable regularity.  The dark
    matter mass profiles show 6\% root-mean-square variation near
    the half-mass radius and the baryonic mass profiles show 17.7\%
    RMS variation (compared to 51\% and 18.5\% respectively for the
    conventional observationally motivated set of scaling constants)
    in spite of the wide range of masses, gas fractions, orbits, and
    feedback recipes employed in these simulations.}
  \label{fig:homology-density-meq}
\end{figure}

Figure \ref{fig:homology-density-meq} shows a different choice of
length and mass scales.  The length scale is \req, the radius at
which the enclosed stellar and dark masses are equal, and the mass
scale is the stellar mass enclosed within \req.  In comparing
to the mass profiles shown in Figure \ref{fig:homology-density}, we
must ensure that the same physical radial range of the galaxies are
probed.  Multiplying the chosen units by a factor common to all
galaxies does not change the result, so we normalize to the \sbc201a-u4
simulation.  For this simulation, $\reff = 0.437\req$ and $\mb =
1.54 \meq$.  

The simulated galaxy merger remnants analysed here are not perfectly
scaling-homologous.  This is clear as soon as one realizes that the
central slopes of the dark matter profiles are different between the G
and \sbc\ series of simulations.  Changing the scale radius can only
shift curves from side-to-side in Figures \ref{fig:homology-density}
and \ref{fig:homology-density-meq}; it cannot change the slopes.  Even
so, the radius and mass scale choice made in Figure
\ref{fig:homology-density-meq} represents an impressively high degree
of regularity among galaxy merger remnants given the variation in mass
ratios, orbits, and gas fractions.  

The total mass density profile for all of the simulations is very
close to an isothermal $\rho \propto r^{-2}$ profile.  This is in good
agreement with recent studies of the mass profiles of elliptical
galaxies using strong and weak gravitational lensing
\citep{gavazzi:07, gavazzi:08, bolton:08-fp}, dynamical studies of the
line-of-sight velocity profiles of giant elliptical galaxies
\citep{gerhard:01}, and studies of the density and temperature
profiles of x-ray gas in elliptical galaxies spanning a wide range in mass
\citep{humphrey:10}. 

Table \ref{tab:variation} lists the RMS variation among all
simulations for each physical quantity considered in this section.  We take the
value of the non-dimensional quantity under consideration at the
appropriate scale radius (either \reff\ or 0.437\req) and compute the
RMS variation of the resulting set of numbers.  This serves as a
indication of the galaxy-to-galaxy variation of, for example, the
density profiles after applying a given set of scaling constants, and
therefore of the utility of that particular set of scaling constants.  

\begin{table}
  \centering
 \begin{minipage}{\columnwidth}
  \caption{Root-mean-square variation of non-dimensional quantities
    among all simulations at the given scale radius.  
The conventional observationally-motivated choice of scaling constants
is used for the column on the left (\reff, \mb, and $\sigma_0$), while
our preferred choice of scaling constants based on the radius at which
the enclosed dark and luminous masses are equal (\req, \meq, and \teq)
is used for the column on the right.  The quantities listed are the
total mass density, the baryonic mass density, the dark matter mass
density, the radial 3D stellar velocity dispersion, the 3D stellar velocity
dispersion in the azimuthal direction, the 3D stellar
velocity dispersion in the $\theta$ direction, the mean streaming
velocity in the azimuthal direction, the
specific kinetic energy of the stellar component, the specific kinetic
energy of the dark matter component, the stellar velocity anisotropy
in the azimuthal direction, and the stellar velocity anisotropy in the
$\theta$ direction.  All of the values
are expressed as a percentage of the mean value for all galaxies
except for the anisotropy $\beta$, where we do not divide by the mean
because anisotropy is not positive definite.}
  \label{tab:variation}
  \begin{tabular}{@{}lr@{}lr@{}l@{}}
  \hline
   Quantity     & \multicolumn{2}{c}{\reff\ based scaling} & \multicolumn{2}{c}{\req\ based scaling} \\ 
\hline
\hline
\rhob+\rhod    & 30&.1\% & 10&.1\% \\
\rhob          & 18&.5\% & 17&.7\% \\
\rhod          & 51&.3\% & 6&.04\% \\
$\sigma_r$     & 26&.9\% & 15&.0\% \\
$\sigma_\phi$   & 37&.6\% & 21&.0\% \\
$\sigma_\theta$ & 40&.8\% & 19&.8\% \\
$v_\phi$       & 188&\% & 176&\% \\
$E_k/m$ (stars)  & 94&.9\% & 11&.0\% \\
$E_k/m$ (dark matter)  & 64&.6\% & 14&.1\% \\
$\beta_\phi$    & 0&.263 & 0&.238 \\
$\beta_\theta$  & 0&.176 & 0&.166 \\
\hline
\end{tabular}
\end{minipage}
\end{table}

\begin{figure}
  \centering
  \includegraphics[width=0.9\columnwidth]{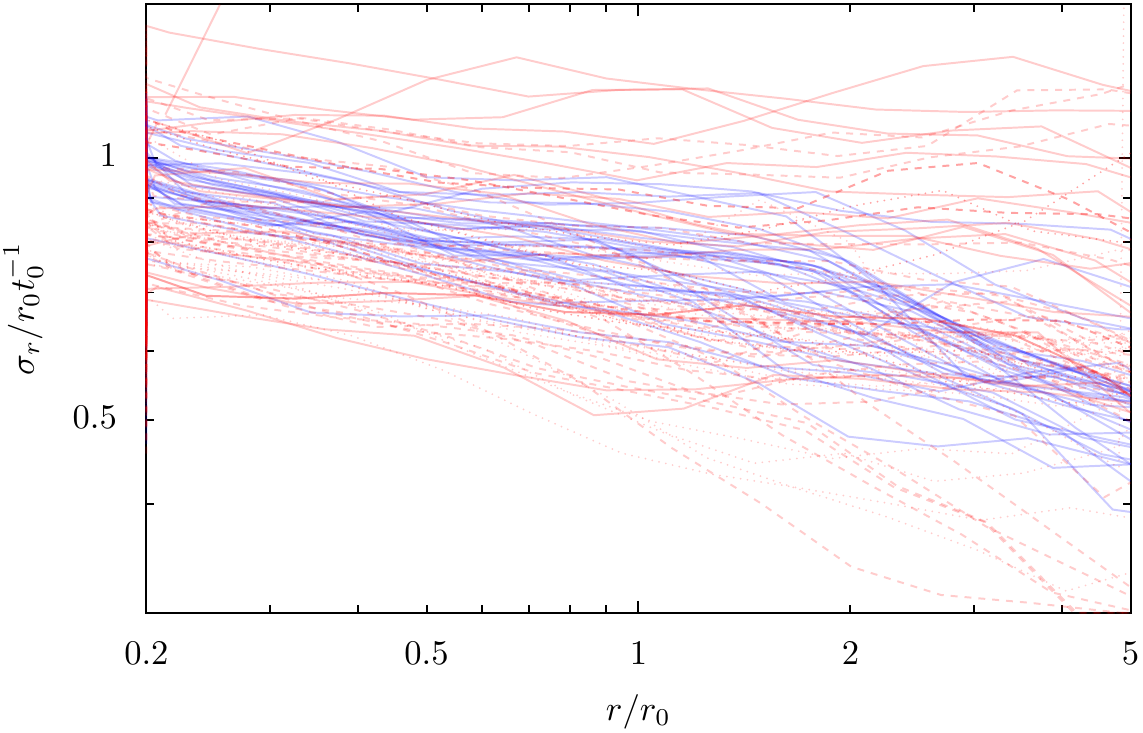}
  \includegraphics[width=0.9\columnwidth]{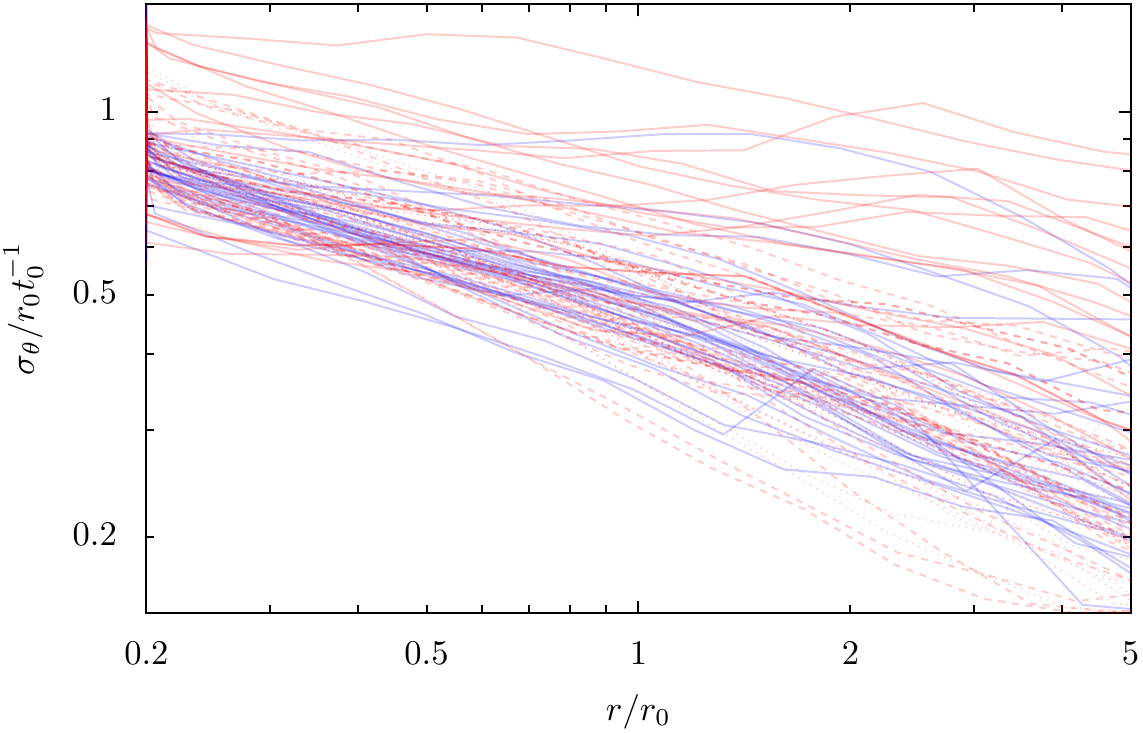}
  \includegraphics[width=0.9\columnwidth]{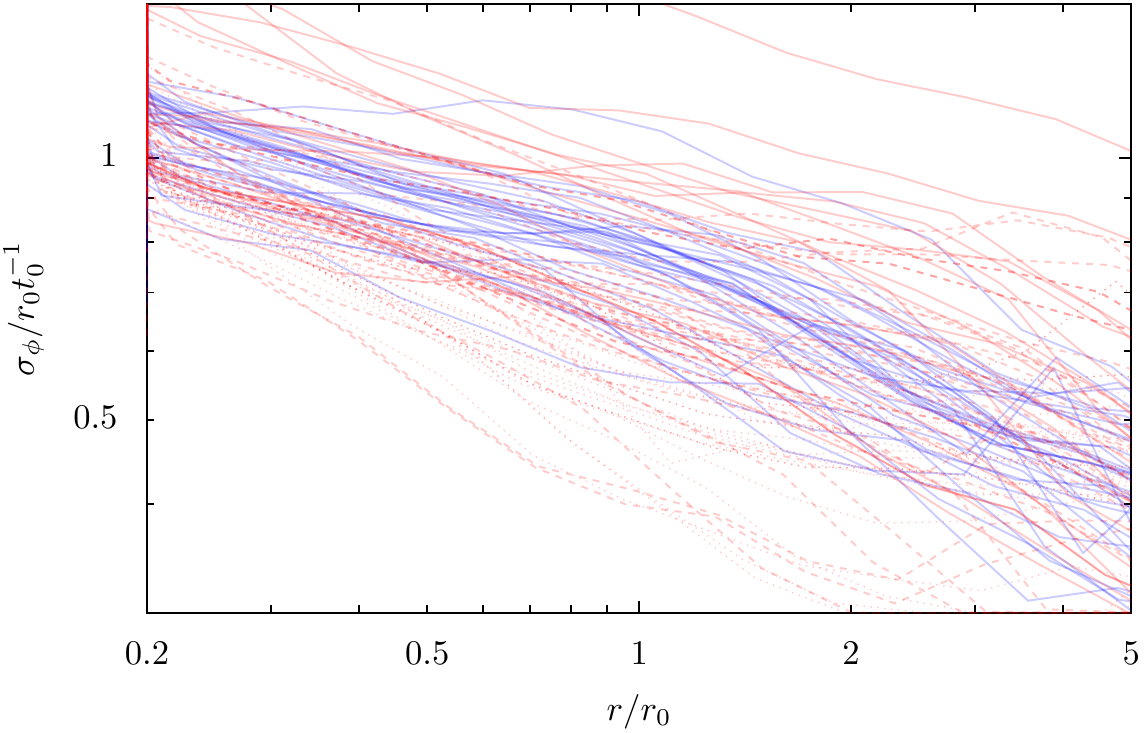}
  \includegraphics[width=0.9\columnwidth]{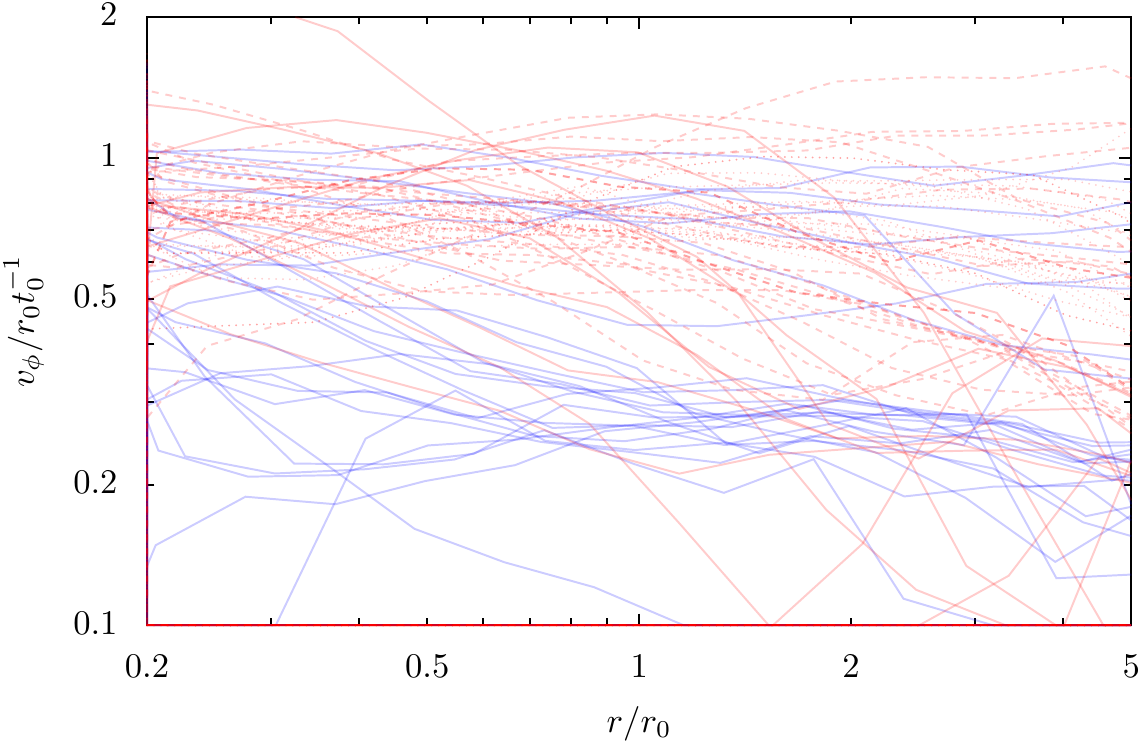}  
  \caption{Dimensionless stellar velocity profiles with scale
    radius and velocity taken to be the usual observationally motivated
    ones: $r_0$ is the is the stellar half-mass radius and $\sigma_0$ is
    the projected velocity dispersion within an aperture of $\reff/8$.
    From the top: 3D radial velocity dispersion, 3D velocity
    dispersion in the $\theta$ direction, 3D velocity
    dispersion in the azimuthal direction, and mean streaming velocity in
    the azimuthal direction.  There is significant diversity in the
    velocity structure of simulated galaxy merger remnants.}
  \label{fig:homology-velocity-stars-re-sigmap}
\end{figure}

\begin{figure}
  \centering
  \includegraphics[width=0.9\columnwidth]{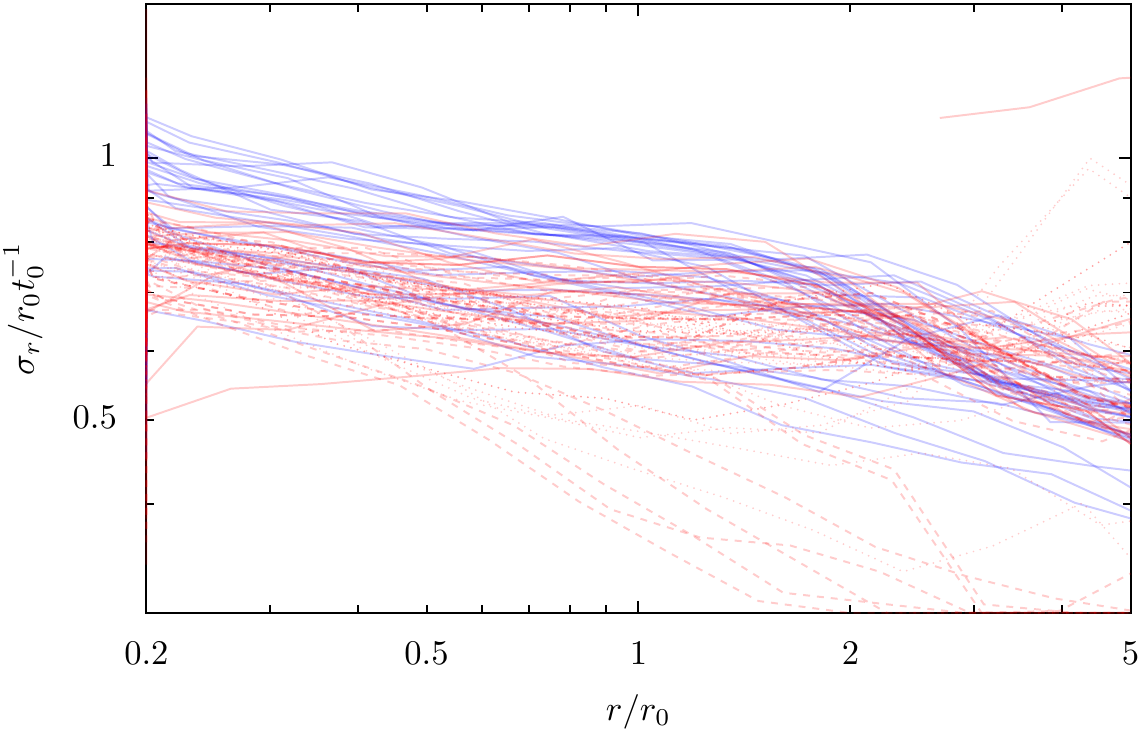}
  \includegraphics[width=0.9\columnwidth]{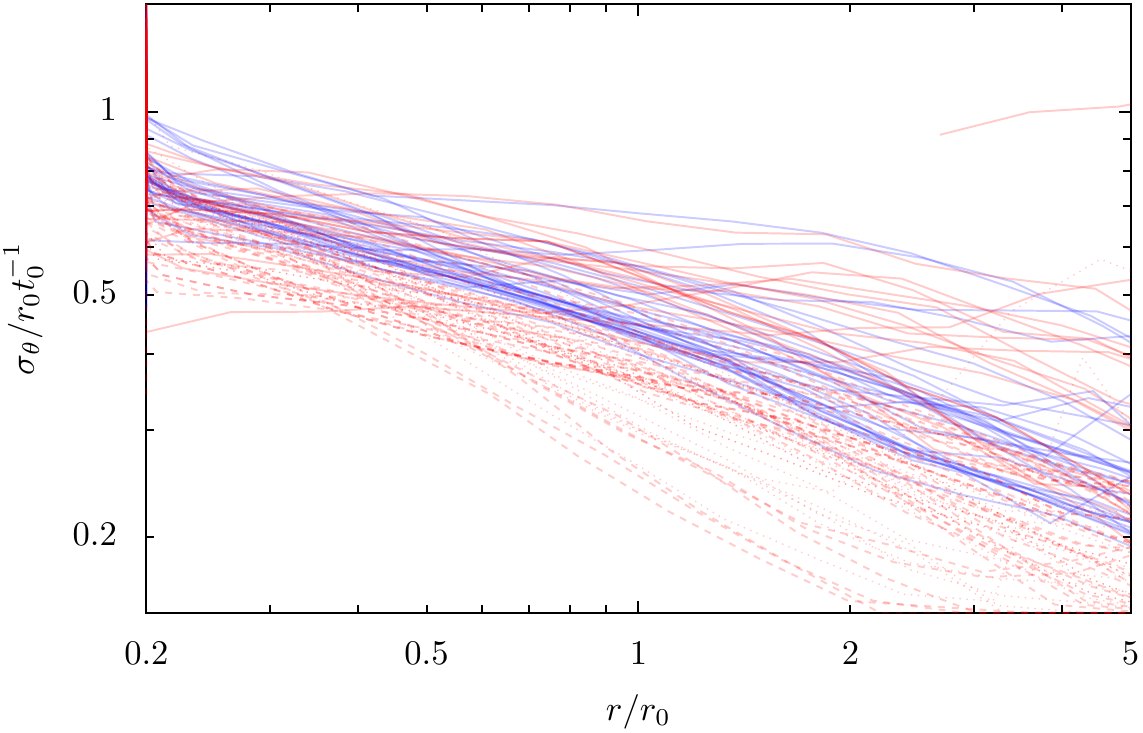}
  \includegraphics[width=0.9\columnwidth]{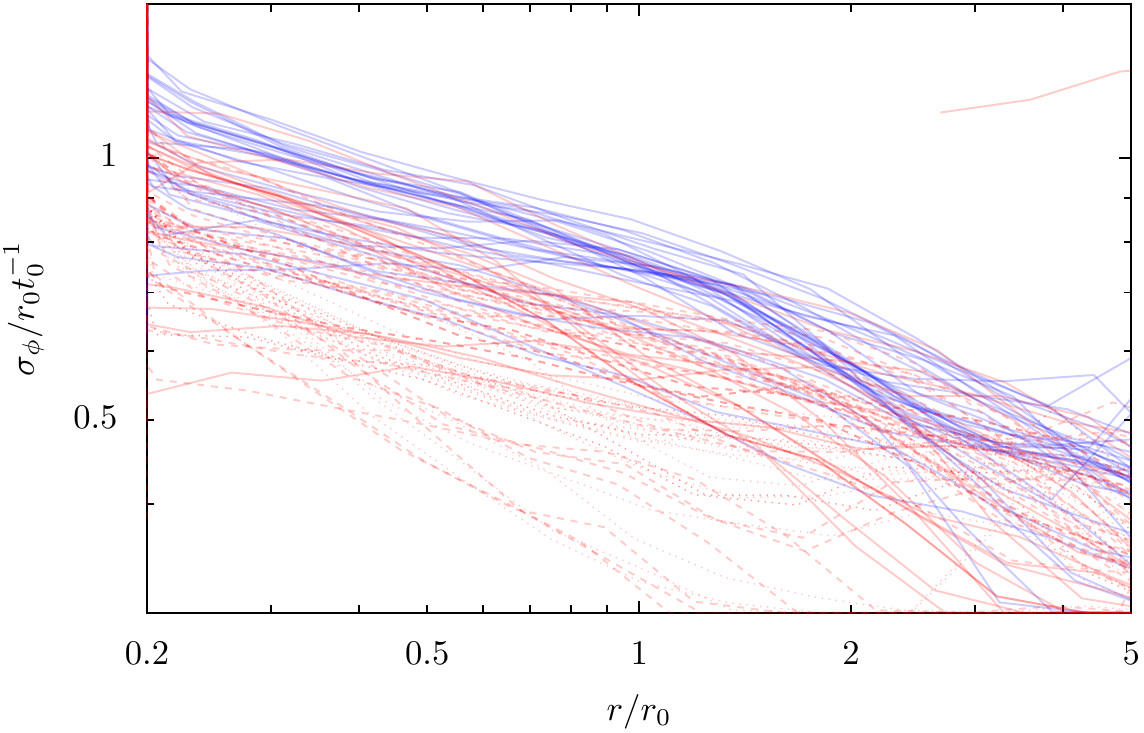}
  \includegraphics[width=0.9\columnwidth]{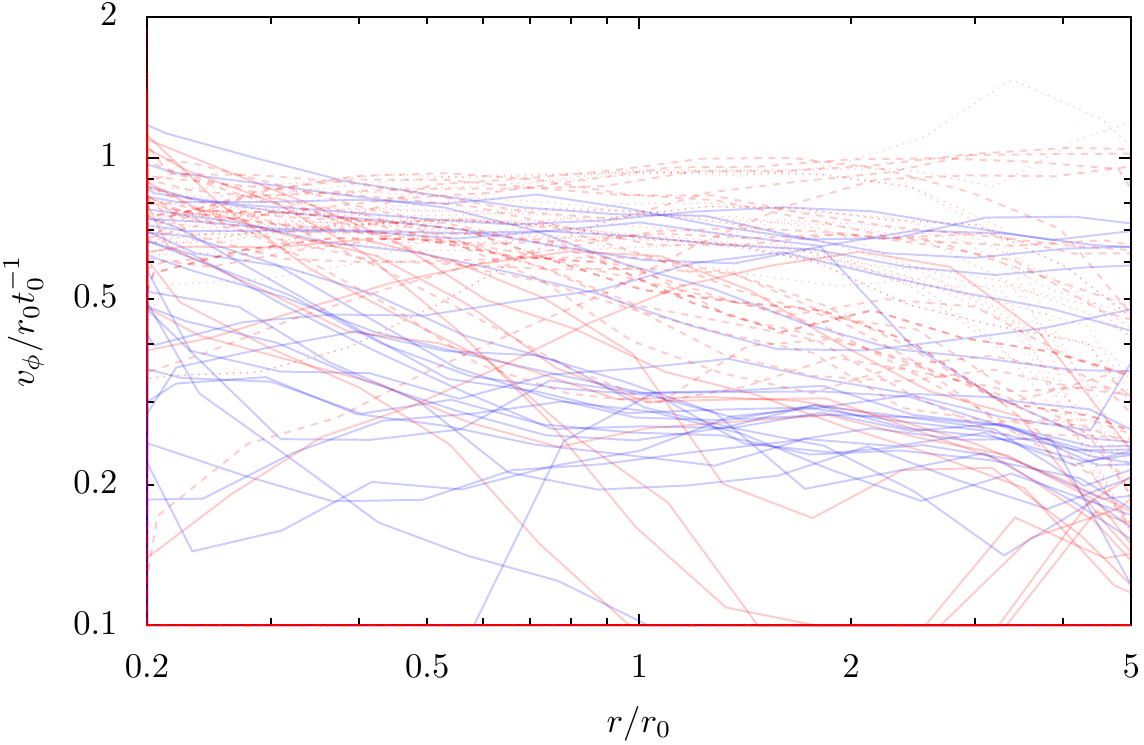}
  \caption{Dimensionless stellar velocity profiles with scale radius
    and time taken to be our preferred choices of \req\ and \teq.  
    This is an improvement over the
    conventional choices for scaling radius and velocity shown in
    Figure \ref{fig:homology-velocity-stars-re-sigmap}: the root-mean-square
    variation of $\sigma_r$, for example, at the scale radius goes
    from 27\% to 15\% in adopting the new set of scaling constants.
    The RMS variations of all quantities shown here are listed in
    Table \ref{tab:variation}.}
  \label{fig:homology-velocity-stars-req-tdyn}
\end{figure}

Figure \ref{fig:homology-velocity-stars-re-sigmap} shows remnant
stellar velocity profiles for the typical choice of scaling constants:
the stellar half-mass radius and the central projected velocity
dispersion within $\reff/8$.  
Figure \ref{fig:homology-velocity-stars-req-tdyn} shows stellar
velocity profiles when scaled by \req\ and the dynamical time at \req.

\begin{figure}
  \centering
  \includegraphics[width=0.9\columnwidth]{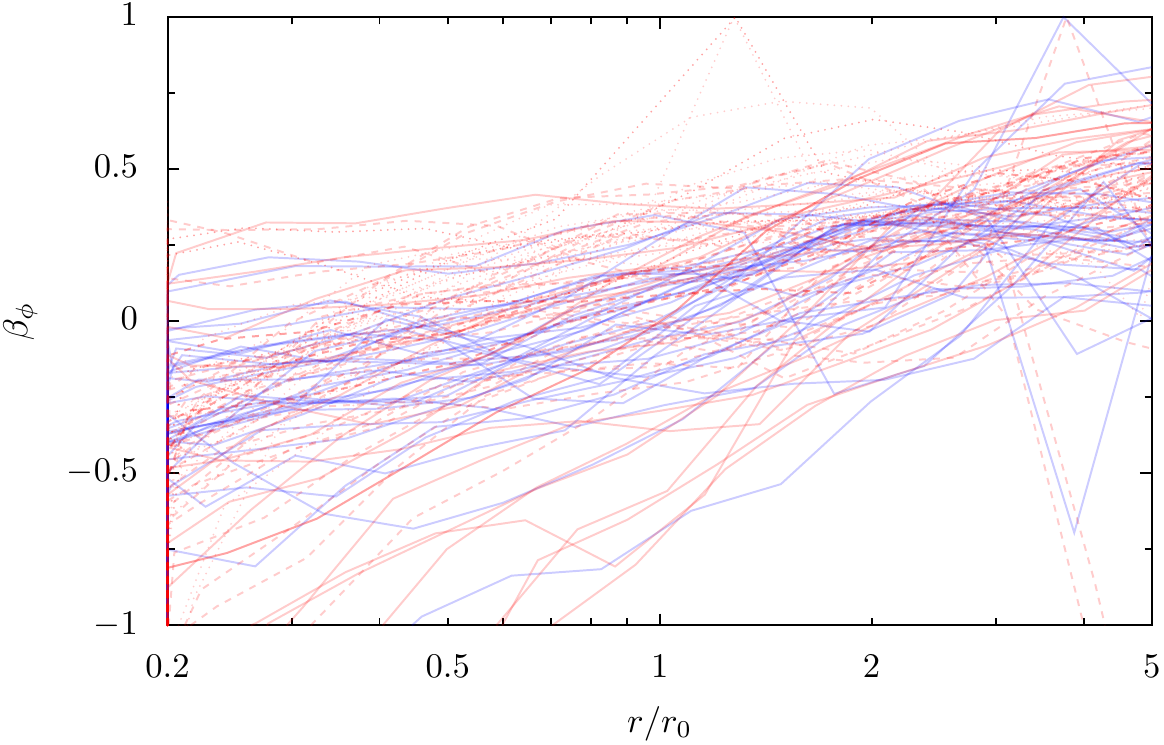}
  \includegraphics[width=0.9\columnwidth]{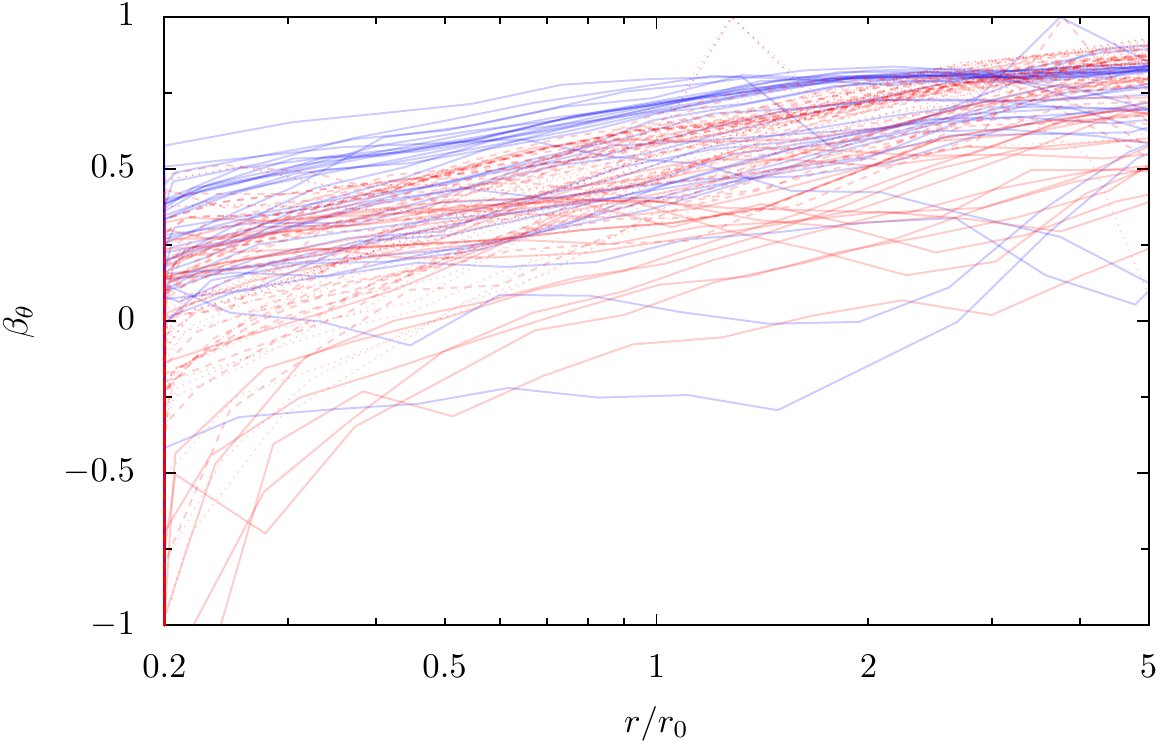}
  \caption{Stellar velocity dispersion anisotropy profiles for the
    usual choice of scale radius: the baryonic half-mass radius.  This
    shows the reason why there is diversity in the velocity profiles
    of galaxy remnants in spite of the regularity in the density
    profiles.  The virial theorem constrains the total kinetic energy
    of the system, but does not determine the {\em direction} of the
    velocities of the stars.  Simulated remnants make use of this
    freedom and have a wide range of anisotropy profiles.  Since the
    velocity anisotropy is dimensionless, changing dimensional scaling
    constants for each galaxy only moves each line horizontally.  The
    shallow slope of the lines means that changing the definition of
    dimensional scaling constants cannot significantly change this
    plot to make the merger remnants more homologous.}
  \label{fig:homology-anisotropy-re}
\end{figure}

The velocity dispersion anisotropy profiles provide a convenient limit
on the extent to which the velocity structure of the simulated
remnants can be brought into a common system.  The anisotropy is
dimensionless, hence the only choice to be made is the length scaling
constant.  The anisotropy $\beta_{\rm a}$ is usually defined for a spherically
symmetric object.  In that case, $\sigma_\theta=\sigma_\phi$ and $v_r
= v_\theta = v_\phi = 0$ The merger remnants presented here are not
spherically symmetric so we define 
\begin{equation}
\beta_\theta = 1-\sigma_\theta^2/\sigma_r^2
\end{equation}
and
\begin{equation}
\beta_\phi = 1-\sigma_\phi^2/\sigma_r^2
\end{equation}
Figure \ref{fig:homology-anisotropy-re} shows both $\beta_\theta$ and $\beta_\phi$.

Virtually all of the remnants show stellar anisotropy profiles that
rise with radius, in some cases to values near 1 (completely radial
velocity dispersion).  This is easily understood in terms of the
dynamical origin of the stars at large radius in merger remnants
\citep{dekel:05}.  Stars that end up at large radius are stars that
were flung out from near the centre of mass when the merging galaxies
underwent their first close pass.  They have low angular momentum
because they started out within approximately one effective radius, 
hence the anisotropy measured at large radius will be very radial.   

The slopes of the anisotropy
profiles are similar from galaxy to galaxy, but they are shallow.
Changing the scale radius slides the curves horizontally, so any scale
radius that would bring all the curves into agreement would be forced
to vary dramatically from galaxy to galaxy.  

We have seen that the density profiles of simulated remnants are
remarkably consistent, but the velocity profiles show significant
variation.  Figure \ref{fig:homology-anisotropy-re} suggests that the
reason for the variation in the velocity profiles is the direction of
the velocities of the stars.  That is, there is a common 
{\em kinetic energy} profile, but the velocity profiles are different because
some remnants are radially anisotropic, some are tangentially anisotropic,
and some are rotating.  

The specific kinetic energy is:
\begin{equation}
  E_k/m = \frac{1}{2}(v_\phi^2 + \sigma_r^2 + \sigma_\theta^2 + \sigma_\phi^2)
\end{equation}
where we have assumed Gaussian velocity distributions and neglected
$v_r$ and $v_\theta$ because the remnants are in steady-state.

\begin{figure}
   \centering
  \includegraphics[width=0.9\columnwidth]{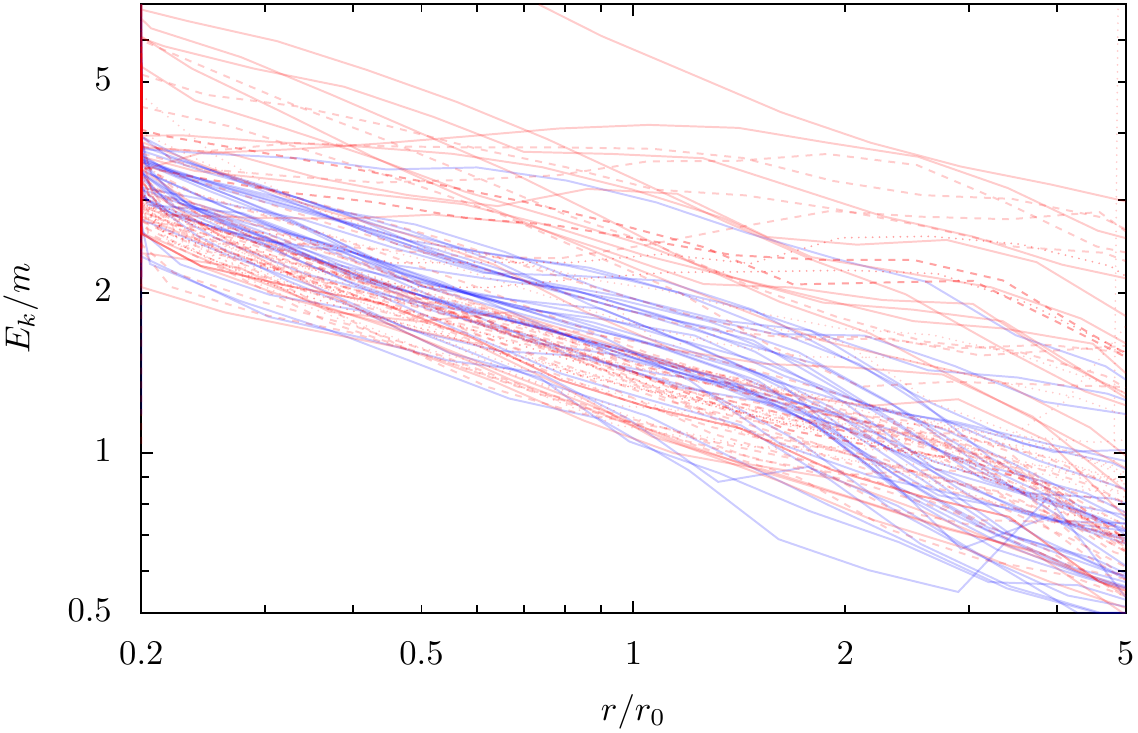}
  \includegraphics[width=0.9\columnwidth]{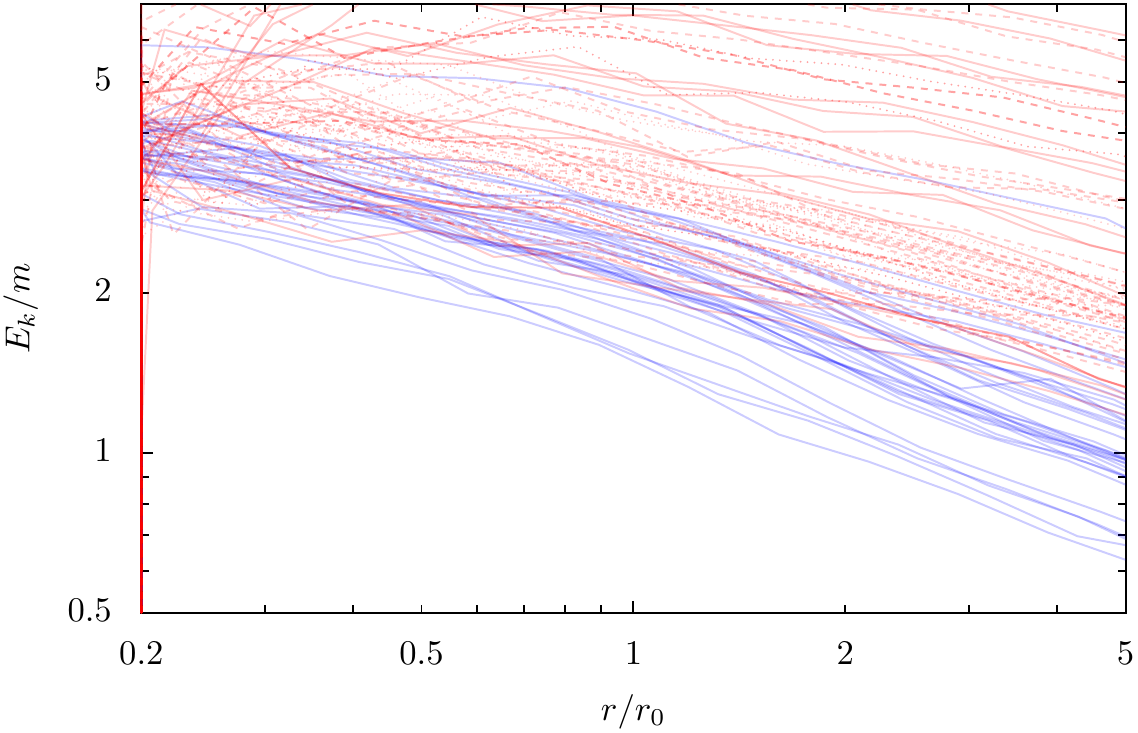}
  \caption{
    Specific kinetic energy profiles with scale constants $r_0$
    equal to the stellar half mass radius and $t_0=r_0/\sigma_0$.  The
    upper panel shows the stellar component while the lower panel
    shows the dark matter component.
    There appears to be great diversity in the kinetic energy profiles
    of galaxy remnants.  }
  \label{fig:homology-ke-1}
\end{figure}

Figure \ref{fig:homology-ke-1} shows kinetic energy profiles using the
stellar half-mass radius and central projected velocity dispersion as
scale constants.  A common kinetic energy profile is not apparent with
this choice of scale constants.

\begin{figure}
  \centering
  \includegraphics[width=0.9\columnwidth]{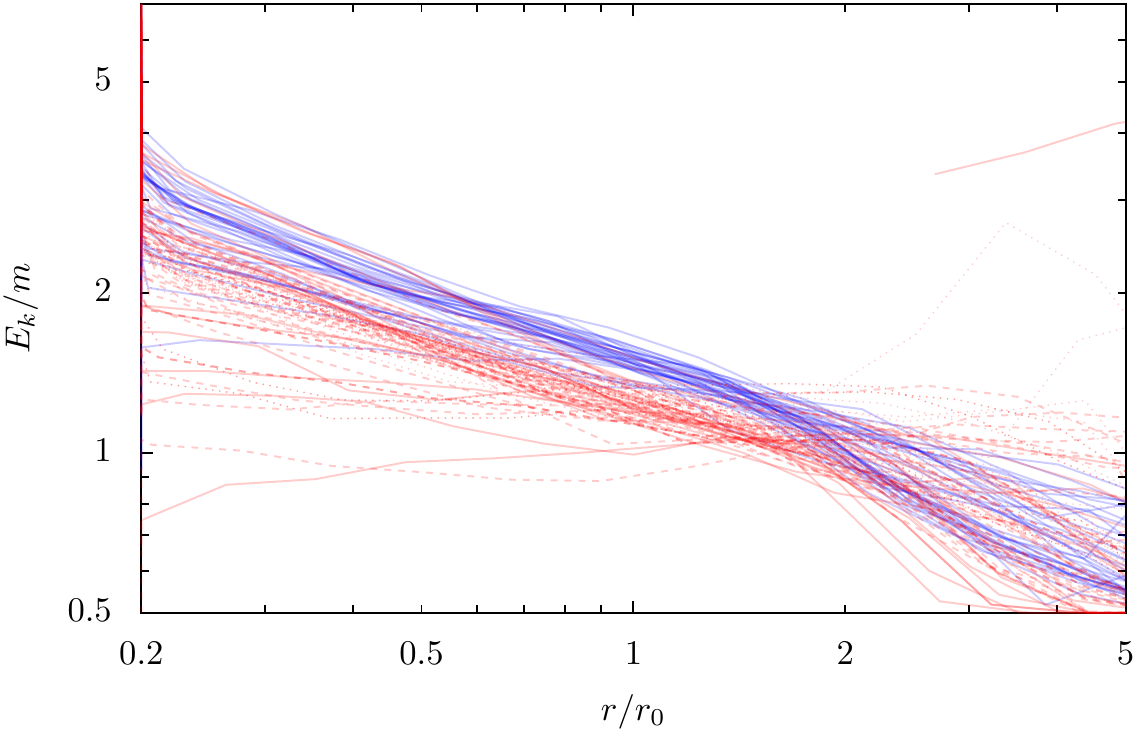}
  \includegraphics[width=0.9\columnwidth]{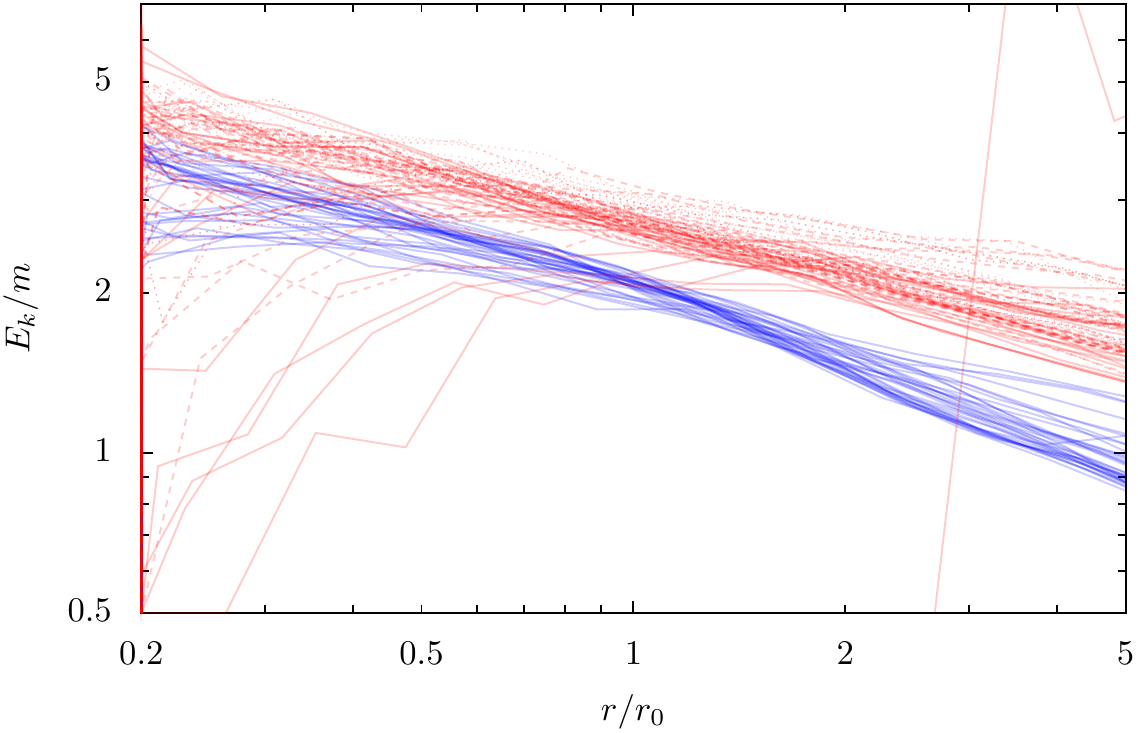}
  \caption{Specific kinetic energy profiles with scale constants taken
    to be our preferred choices of \req\ and the dynamical time at \req.
    The upper panel shows the stellar component and the lower panel
    shows the dark matter component.  These kinetic energy profiles
    show very consistent behaviour given the great diversity of the
    origins of the remnants.}
  \label{fig:homology-ke-3}
\end{figure}

Figure \ref{fig:homology-ke-3} shows the dimensionless kinetic
energy scaled by the mass-equality radius and the dynamical time
at the mass-equality radius.  The common energy structure of the merger
remnants is apparent.  In the case of the \sbc\ simulations there
is as little as 3\% RMS variation at the scale radius.  
There is again an offset between the two sets of
simulations but nevertheless the lack of variation between simulations
of a given set is remarkable.

It is clear that dark matter plays a significant but sub-dominant role
in the dynamical equilibrium of the luminous parts of galaxies, and 
that the baryonic and dark matter components of galaxies
scale very differently with radius.  Observational and theoretical
studies to date have almost exclusively used the classic
observationally accessible half-light radius as the relevant scaling
parameter to compare galaxies of different sizes and masses.  However,
the value of this parameter involves only the luminous matter.  It is
therefore quite plausible that adopting a different scaling radius
will lead to different conclusions about homology: a set of galaxies
(real or simulated) may be closer to homology when using $\req$ as the
scaling parameter.

Unfortunately it is very difficult to determine what conclusions other
studies would have reached if they had used $\req$ instead of $\reff$
without access to the full density and light profiles of the objects
that were used in the study.  One may be able to make reasonable
extrapolations of the profiles from $\reff$ to $\req$, but this would
not test the hypothesis.  It is the conclusion of the present study that
sets of galaxies may look more homologous or less homologous depending
on the scaling radius that one uses, so making an assumption about the
mass profiles in order to translate results from one radius to another
would not test the hypothesis.  One needs access to the full mass
profile (in the case of simulated galaxies) or light profile (in the
case of observational studies).  On this point we hope that other
observers and theorists use the relatively simple procedure of scaling
by $\req$ to see, first, if it makes their results easier to
understand, and, second, if the results of our present study hold for
broader sets of simulations and observations.  We believe that
adopting $\req$ rather than $\reff$ would result in a significant gain
in understanding.  

\section[]{Observational Implications}
\label{sec:observations}

As we have already mentioned, it is unfortunately the case that what
we find to be the best definitions for the dimensional scaling
constants that make simulated merger remnants close to homologous are
not directly observable.  In this section we define a plausible model
for the density profiles of baryonic and dark matter in a two
component galaxy and find the relationship between our preferred
dimensional scaling constants and the observationally accessible
values.  That is, we find $\req/\reff$ and $\meq/\mb$ for a range of
plausible mass models.

It is possible to plot values of $\req$ and $\meq$ for our simulated
galaxy merger remnants but there are several problems with this idea.
Most importantly, anyone who wished to use these values of $\req$ and
$\meq$ to interpret observations would automatically be subject to all
of the uncertainties and deficiencies of numerical simulations, for
example in the star formation law to name one issue among many.
Second, because our galaxies {\em are} so close to homologous when
$\req$ is used as a scaling parameter, the values of $\req/\reff$ and
$\meq/\mb$ are nearly constant.  Third, this particular set of
simulated merger remnants do not span the full space of
observationally relevant parameters, in the S\'ersic index for example.  

Therefore we seek a simple, easily understood, widely used, and
observationally relevant mass model for which we can compute $\req$
and $\meq$.  It is easy to understand the assumptions that enter such a
model and, importantly, the model is not dependent on the details of
our simulations even though the motivation to compute $\req$ and
$\meq$ arose from analysis of our simulations.

It is a relatively simple matter to calculate $\req$ and $\meq$ for
any given combination of dark matter and baryonic mass profiles.  We
provide plots and fitting formulas for the most common cases in the
hope that these formulas may be used in both observational and
theoretical efforts to understand the scaling relations of galaxies.
It is the conclusion of this study that different choices of the
scaling parameters can lead to different conclusions about homology,
so this is potentially an important point for both theoretical and
observational studies of galaxy scaling relations.  

For the baryonic component of our mass model, we use deprojected
S\'ersic profiles \citep{sersic:68}:
\begin{equation}
  I(r_p) = I_0 \exp\left[-\left(\frac{r_p}{r_s}\right)^{1/n} \right]
  \label{eq:sersic}
\end{equation}
where $I$ is the surface brightness, $r_p$ is the projected radius,
$I_0$ is the central surface brightness, and $r_s$ is a scale radius
chosen to ensure that the half-light radius is equal to some desired
value.  The S\'ersic profile is defined for the projected surface
brightness of a galaxy image.  Approximations to the three-dimensional
luminosity density that lead to the S\'ersic profile under the
assumption of spherical symmetry have been available for some time
\citep{prugniel:97, lima-neto:99, trujillo:02,mamon:05}.  Exact
analytic deprojections of the S\'ersic profile for {\em integer
  values} of the S\'ersic index $n$ have also been available for some
time \citep{mazure:02}.  Recently \citet{baes:10} have generalized the
\citet{mazure:02} result for rational values of the S\'ersic index, a
very significant step forward for practical applications of the
formulas since rational numbers can approximate any desired value
arbitrarily well.  These involve special functions, but numerical
implementations of the necessary functions are available.
\citet{baes:10} also provide formulas for the case where the S\'ersic
index is real-valued rather than rational, but numerical
implementations of the required special functions are not readily available.
We use the
\citet{baes:10} formulas for exact deprojections of the S\'ersic
profile.

We allow two possibilities for the dark matter halo.  The first is the
familiar NFW formula \citep{navarro:96}:
\begin{equation}
\rho(r) = \frac{\rhonfw}{x(1+x)^2}  
\end{equation}
where $\rho$ is the mass density, $x$ = $r/\rnfw$, $\rnfw$
is the scale radius where the profile switches from a logarithmic
slope of -1 to -3, and $\rhonfw$ is a constant that sets the dark
matter fraction of the galaxy as a function of radius.

There is good evidence from strong and weak gravitational lensing
studies that the total mass density of elliptical galaxies is
remarkably close to that of a singular isothermal sphere (SIS) over a
large range in radii \citep{gavazzi:07, gavazzi:08, bolton:08-fp}.
Therefore the second possibility for the dark matter halo is the mass
density such that the {\em total} mass density is given by:
\begin{equation}
  \rhot(r) = \frac{s^2}{2 \pi G r^2}
\end{equation}
where $s$ parametrizes the depth of the potential well.  

\begin{figure}
  \centering
  \includegraphics[width=0.9\columnwidth]{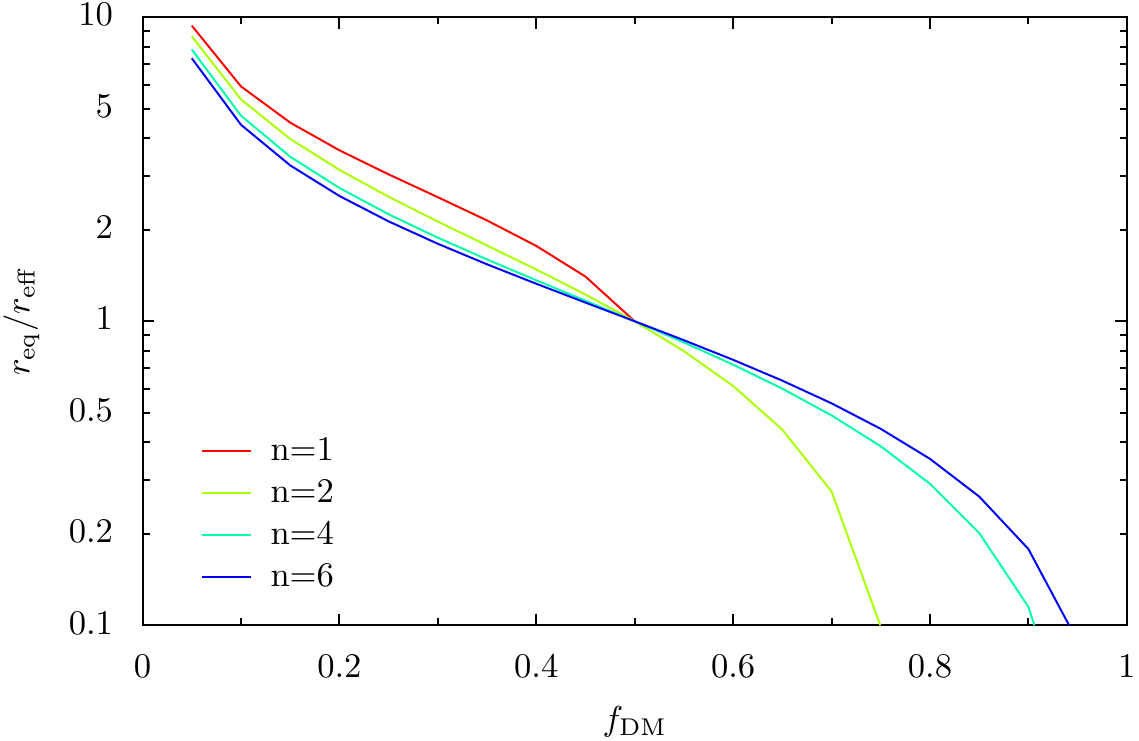}
  \includegraphics[width=0.9\columnwidth]{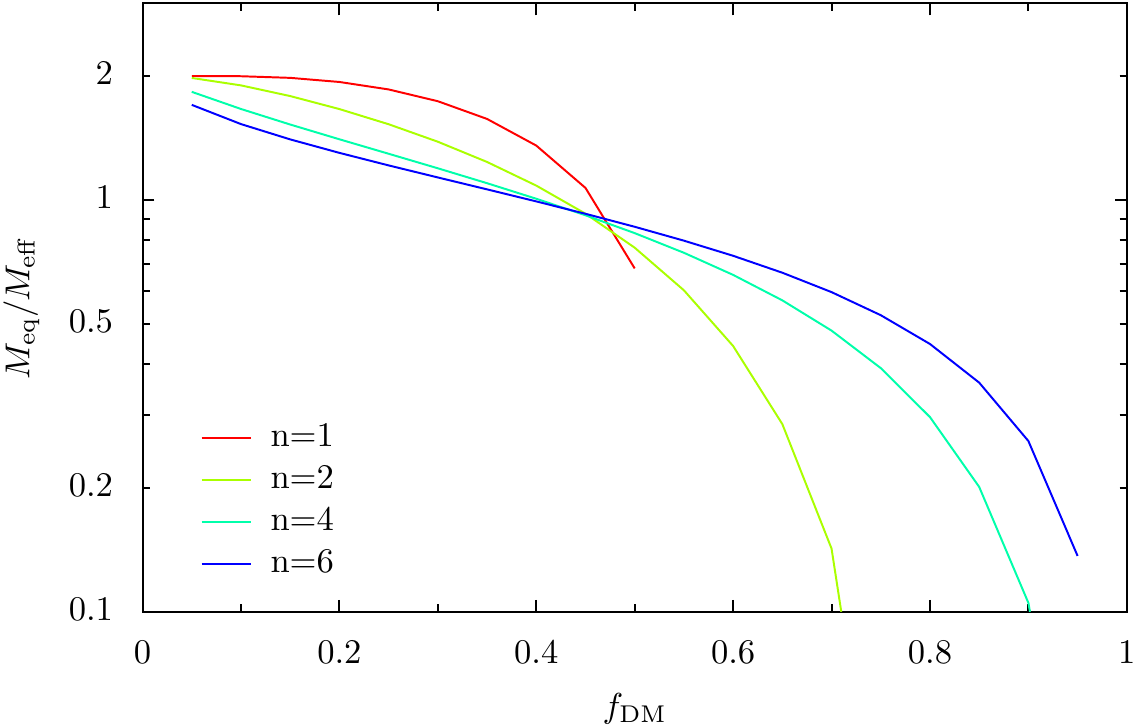}
  \caption{The relationship between our preferred dimensional scaling
    constants and directly observable quantities for a variety of mass
    models.  The upper panel shows \req/\reff\ while the
    lower panel shows \meq/\meff, both as a function of the 3D dark
    matter fraction within the projected baryonic half-light radius.
    The parameter $n$ is defined in equation \ref{eq:sersic}
    and gives the shape of the baryonic density profile.  The dark
    matter fraction within one baryonic effective radius changes as a
    result of changes to the NFW density parameter $\rhonfw$ for a
    fixed baryonic model.  
    The dominant consideration is the dark matter fraction, while the
    shape of the baryonic mass profile has a smaller effect. }
  \label{fig:observables-nfw}
\end{figure}

Figure \ref{fig:observables-nfw} gives $\req/\reff$ and $\meq/\mb$ for
a range of models assuming an NFW halo parametrized by the shape of
the baryonic mass profile and the dark matter fraction within one
baryonic effective radius.  For a fixed baryonic mass and effective
radius, the dark matter fraction within one baryonic projected
half-light radius is adjusted by changing the NFW density parameter
$\rhonfw$.  The results depend on the ratio of the $\rnfw/\reff$ which
indicates whether the density profile of the dark matter halo changes
logarithmic slope near the baryonic component or far outside of it.
The dependence is weak as long as $\rnfw/\reff$ is sufficiently large,
with the values of $\req$ and $\meq$ approaching constant values.  The
plots and fitting formulas presented here assume $\rnfw/\reff=20$.
The values of $\req$ are $\sim 10\%$ smaller for $\rnfw/\reff=\infty$
and $\sim 10\%$ larger for $\rnfw/\reff$ = 10, with little dependence
on S\'ersic index or central dark matter fraction.  When $\rnfw/\reff$
approaches unity, $\req$ depends sensitively on the ratio, but that
corresponds to halo concentrations $C = r_{\rm virial}/\rnfw$
approaching 100, far outside the range predicted by simulations for
main haloes \citep{bullock:01}, although tidal stripping can lead to
very large values of the concentration for sub-haloes \citep{diemand:08}.

The value of $\req$ for the case of an NFW halo may be approximated
as:
\begin{eqnarray}
  \log_{10}{\frac{\req}{\reff}} &=& \log_{10}f_1(f_{\rm DM}) +
  f_2(f_{\rm DM}) \log_{10}(n/3) \nonumber \\ 
& & + f_3(f_{\rm DM}) \log^2_{10}(n/3)
  \label{eq:fit-start}
\end{eqnarray}
\begin{equation}
  f_1(x) = k_1 (x/k_2)^{k_3} 10^{-k_4(x/k_2)^{k_5}}
\end{equation}
\begin{equation}
  f_2(x) = k_6 + k_7(x/0.5) + k_8(x/0.5)^2 + k_9(x/0.5)^3
\end{equation}
\begin{equation}
  f_3(x) = k_{10} + k_{11}(x/0.5) + k_{12}(x/0.5)^2 + k_{13}(x/0.5)^3
  \label{eq:fit-end}
\end{equation}
where $n$ is the S\'ersic index, $f_{\rm DM}$ is the 3D fraction of
dark matter to total matter within the projected baryonic half-light
radius, and the constants $k_1$ through $k_{13}$ are given in Table
\ref{tab:nfw-constants}.  This approximation is accurate to 8\% for
S\'ersic indices between 1 and 8 and and dark matter fractions
between 5\% and 70\%.  This formula gives the values for
$\rnfw/\reff=20$ and is subject to modification described above.

\begin{table}
  \centering
 \begin{minipage}{140mm}
  \caption{Fitting formula constants for NFW haloes.}
  \label{tab:nfw-constants}
  \begin{tabular}{@{}lr@{}l@{}}
  \hline
   Name     & \multicolumn{2}{c}{Value} \\ 
\hline
\hline
$k_1$ & 1&.183 \\
$k_2$ & 0&.7252 \\
$k_3$ & 0&.7241 \\
$k_4$ & 0&.5476 \\
$k_5$ & 2&.752  \\
$k_6$ & -0&.1582 \\
$k_7$ & 0&.09714\\
$k_8$ & -0&.6602\\
$k_9$ & 0&.7261 \\
$k_{10}$ & -0&.05721 \\
$k_{11}$ & -0&.005238 \\
$k_{12}$ & 1&.370\\
$k_{13}$ & -1&.319\\
\hline
\end{tabular}
\end{minipage}
\end{table}

\begin{figure}
  \centering
  \includegraphics[width=0.9\columnwidth]{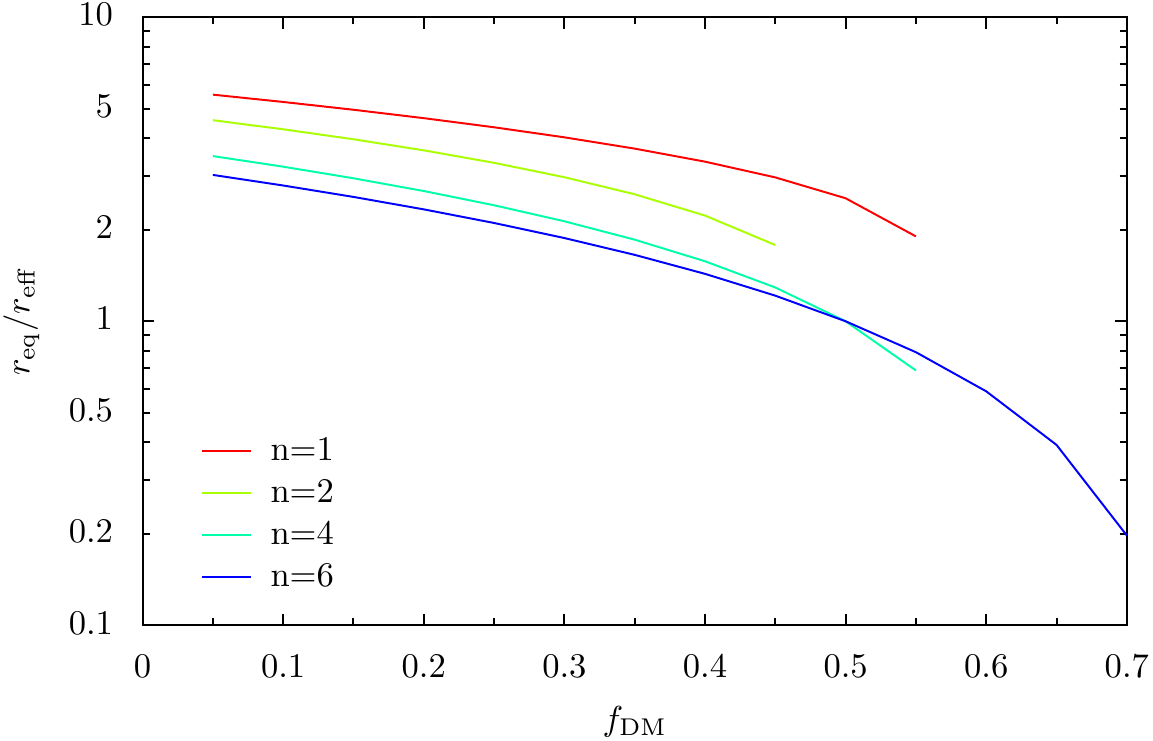}
  \includegraphics[width=0.9\columnwidth]{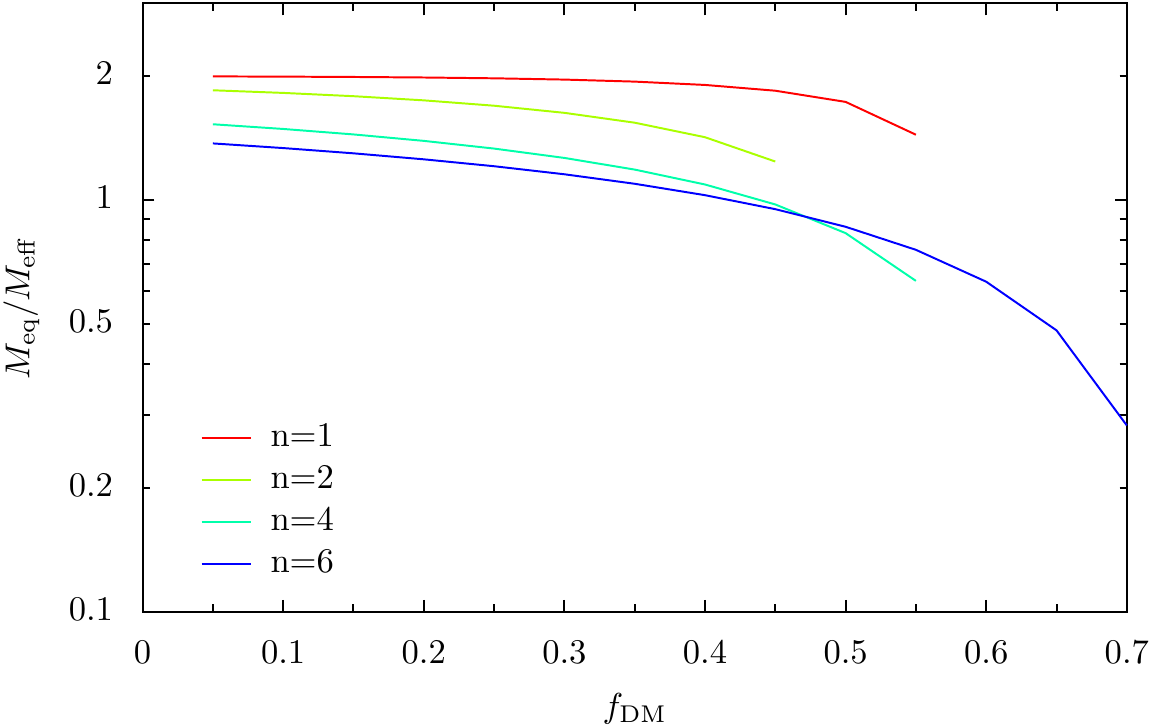}
  \caption{The relationship between our preferred dimensional scaling
    constants and directly observable quantities for a variety of mass
    models.  The upper panel shows \req/\reff\ while the lower panel
    shows \meq/\meff, both as a function of the 3D dark matter
    fraction within the projected baryonic half-light radius.  The
    parameter $n$ is defined in equation \ref{eq:sersic} and gives the
    shape of the baryonic density profile.  The dark matter fraction
    within one baryonic effective radius changes as a result of
    changes to the SIS density parameter $s$ for a fixed baryonic
    model.  The dominant consideration is the dark matter fraction,
    while the shape of the baryonic mass profile has a smaller
    effect. }
  \label{fig:observables-sis}
\end{figure}

Figure \ref{fig:observables-sis} gives $\req/\reff$ and $\meq/\mb$ for
different central dark matter fractions under the assumption that the
total density profile is that of a singular isothermal sphere.  For a
fixed baryonic mass and effective radius, the dark matter fraction
within one baryonic projected half-light radius is adjusted by
changing the SIS parameter $s$.

In this case $\req$ and $\meq$ may be approximated by the same set of
equations as for the case of an NFW halo: equations \ref{eq:fit-start}
through \ref{eq:fit-end}.  Constants for the equations are given in
Table \ref{tab:sis-constants}.  This approximation is accurate to 5\%
for S\'ersic indices between 1 and 8 and dark matter fractions
between 0.05 and 0.7, except near the transition between the two sets
of fitting formula constants at $f_{\rm DM} = 0.55$, where the
approximation is only good to 20\%.  

\begin{table}
  \centering
 \begin{minipage}{140mm}
  \caption{Fitting formula constants for SIS haloes.}
  \label{tab:sis-constants}
  \begin{tabular}{@{}lr@{}lr@{}l@{}}
  \hline
   Name     & \multicolumn{2}{c}{Value for $f_{\rm DM} < 0.55$} & \multicolumn{2}{c}{Value for $f_{\rm DM} >= 0.55$}  \\ 
\hline
 \hline
$k_1$ &  2&.546 & 2&.546 \\
$k_2$ & 0&.6363 & 0&.6363 \\
$k_3$ & 0&.1754 & 0&.1754 \\
$k_4$ &  1&.097 &  1&.097 \\
$k_5$ &  3&.779 &  3&.779 \\
$k_6$ &  -0&.3725 & -43&.72 \\
$k_7$ & 0&.2549 & 38&.30 \\
$k_8$ & -1&.007 & 0 \\
$k_9$ & 0&.7579 & 0 \\
$k_{10}$ & -0&.1034 &  47&.96 \\
$k_{11}$ & 0&.8789 & -42&.32 \\
$k_{12}$ & -2&.0134 & 0 \\
$k_{13}$ & 2&.094 & 0 \\
\hline
\end{tabular}
\end{minipage}
\end{table}

\section[]{Conclusions}
\label{sec:conclusions}

We have considered the exact meaning of the concept of homology in the
context of multi-component galaxies.  A particularly useful
formulation of the concept is termed {\em scaling-homology}.  Using
this definition, homology means that all galaxies are the same when
scaled using the correct choice of length, mass, and time units, where
the units are allowed to change from galaxy to galaxy.  The three
dimensional constants serve to fully specify a galaxy once the
`master' distribution function that describes all galaxies is
known.  

We used a set of hydrodynamical binary galaxy merger simulations
spanning a range of masses, mass ratios, orbits, gas fractions,
and supernova feedback prescriptions (but no AGN feedback) to
determine the definition of dimensional scaling constants that served
to make the remnants as close as possible to homologous.  We found
that the best definition of length, mass, and time units are $\req$,
the radius within which dark and baryonic masses are equal, the mass
within $\req$, and the dynamical time measured at $\req$.  This is
in contrast to the usual observationally motivated definitions of
dimensional scaling constants: the baryonic half-light radius, the
baryonic mass, and the central aperture velocity dispersion.  

Simulated gas-rich binary galaxy merger remnants show remarkably
regular structure when the correct scale constants are used to plot
dimensionless density and kinetic energy profiles.  The stellar
anisotropy is essentially the sole source of the variation in the
kinematic structure of these simulated merger remnants.  Dark halo
concentration and progenitor gas fraction contribute to a systematic
difference between the \sbc\ and \g\ series simulations, but within
each set of simulations the structure is remarkably consistent.

We emphasize that it is not {\em incorrect} to phrase questions about
homology in terms of $\reff$.  Previous theoretical and observational
studies that have done so have correctly asked and answered well-posed
questions about galaxy scaling relations.  However, we find that it is
more informative to phrase questions about homology in terms of $\req$
rather than $\reff$, at least for our set of numerical simulations.
Using $\reff$, we would have concluded that our merger remnants are
not homologous owing to a systematic change in the dark matter
fraction within $\reff$.  However, if we instead scale by $\req$, we
find that the merger remnants appear significantly more uniform.  They
are not perfectly homologous, but they are closer to homology than we
would conclude based on $\reff$.  We are eager to learn whether the
same holds true for numerical simulations by other groups and whether
it holds true for actual galaxies.

In order to facilitate the use of these scaling constants to analyse
actual galaxies, we calculated the ratios $\req/\reff$ and
$\meq/\mb$ for a set of observationally plausible two-component
galaxy models and provided fitting formulas for the results.  

The fact that \req\ plays such a prominent role in these scaling
constants immediately recalls the disc-halo conspiracy arising from
the study of spiral galaxy rotation curves \citep{bahcall:85,
  burstein:85, kent:87}.  \citet{blumenthal:86} argued that such a
conspiracy will arise naturally if the dissipational collapse of
baryons is limited by their initial angular momentum and if the baryon
fraction is roughly equal to the spin parameter $\lambda = J E^{1/2}
G^{-1} M^{-5/2}$.  The result presented here appears to be another manifestation of
this phenomenon.  If the baryons have very little angular momentum,
they will contract to the point where they are completely
self-gravitating and the dark matter does not play a role in the
dynamics, in which case \reff\ and not \req\ is the important
radius.  On the other hand, if the baryons have too much angular
momentum, they will not contract enough and the more massive dark
matter halo will dominate the dynamics while the baryons function more
as a tracer population.  Simulated galaxy merger remnants are nearly
homologous when \req\ rather than \reff\ is used as the
dimensional scaling constant, indicating that galaxies are in a
`Goldilocks' state where both baryons and dark matter play important
dynamical roles.  This is likely because of the initial balance
between the baryon fraction and the spin parameter of dark matter
haloes. 

\section*{Acknowledgments}

We thank Michele Cappellari for useful discussions.  GSN acknowledges
the support of the Princeton University Council on Science and
Technology.  PJ acknowledges support by a grant from the W.M. Keck
Foundation and by program HST-AR-11758, provided by NASA through a
grant from the Space Telescopes Research Institute, which is operated
by the Association of Universities for Research in Astronomy,
Incorporated, under NASA contract NAS5-26555.

\bsp


\label{lastpage}

\end{document}